\documentclass[onecolumn,journal,12pt]{IEEEtran}

\usepackage[cmex10]{amsmath}
\usepackage[dvipdfmx]{graphicx}
\usepackage{array}
\usepackage{mdwmath}
\usepackage{mdwtab}
\usepackage{amssymb}
\usepackage{multirow}
\newtheorem{theorem}{Theorem}
\newtheorem{lemma}{Lemma}

\newtheorem{corollary}{Corollary}
\newtheorem{remark}{Remark}
\newtheorem{proposition}{Proposition}
\newcommand{\qed}{\hfill \IEEEQED}

\DeclareMathAlphabet{\bm}{OML}{cmm}{b}{it}
\newcommand{\rom}[1]{\mathrm{#1}}

\allowdisplaybreaks[2]

\def\Label#1{\label{#1}\ [\ \text{\protect{#1}}\ ]\ }
\def\Label{\label}

\def\PF#1{\noindent{\sf #1}:\quad}
\newcommand{\bZ}{\mathbb{Z}}

\begin{document}
\title{Semi-Finite Length Analysis for 
Information Theoretic Tasks}
\author{
 \IEEEauthorblockN{Masahito Hayashi}\\
  \IEEEauthorblockA{Graduate School of Mathematics, Nagoya University \\
Shenzhen Institute for Quantum Science and Engineering, Southern University of Science and Technology\\
Centre for Quantum Technologies, National University of Singapore \\
 Email: {masahito@math.nagoya-u.ac.jp} }
}


\maketitle

\begin{abstract}
We focus on the optimal value for various information-theoretical tasks.
There are several studies for the asymptotic expansion for 
these optimal values up to the order $\sqrt{n}$ or $\log n$.
However, these expansions have errors of the order 
$o(\sqrt{n})$ or $o(\log n)$, which does not goes to zero asymptotically.
To resolve this problem, 
we derive the asymptotic expansion up to the constant order 
for upper and lower bounds of these optimal values.
While the expansions of upper and lower bonds  do not match,
they clarify the ranges of these optimal values, whose errors go to zero asymptotically.
%\boldmath
\end{abstract}


% \fi
\IEEEpeerreviewmaketitle

\section{Introduction}
Recently, second order analysis and finite-length analysis attract much attention \cite{strassen,Kontoyiannis,Kontoyiannis2,H08,hayashi:09,polyanskiy:10}.
However, there is a gap between these two analyses as follows.
To see this difference, we focus on secure random number generation when a partial information is leaked to the third party \cite{Mau93,AC93}.
It is very useful to calculate 
the maximum size $N_n^\varepsilon$ of secure keys under the constraint 
that the secrecy measure is less than $\varepsilon$
when $n$ outcomes are generated according to 
the independent and identical distribution of a certain distribution.
However, its calculation amount is extremely large so that we cannot calculate it in a realistic time 
with a practical length $n$.
Instead of this evaluation, in second order analysis, we derive asymptotic expansion of $\log N_n^\varepsilon$ up to the order $\sqrt{n}$ as
$A_1 n + A_{2,\varepsilon}\sqrt{n}+ o(\sqrt{ n}) (\hbox{or } O(\log n))
$ \cite{Kontoyiannis,Kontoyiannis2,hayashi:09,polyanskiy:10,epsilon}.
Hence, $A_1 n + A_{2,\varepsilon}\sqrt{n}$ can be regarded as an approximation of $\log N_n^\varepsilon$.
However, since the error behaves as $o(\sqrt{ n})
 (\hbox{or } O(\log n))$, it is difficult to evaluate 
the error of this approximation.
Hence, even when we draw the graph of this approximation, 
it is not easy to identify the true value of $\log N_n^\varepsilon$ in the graph.
In the third order analysis, we derive its asymptotic expansion up to the order $\log n$ 
like $A_1 n + A_{2,\varepsilon}\sqrt{n}+ A_3 \log n + O(1)$ \cite{Polyanskiy,TT1,TT2,Moulin}.
However, it is still difficult to evaluate the error of the approximation because it behaves as 
an unknown constant.
Instead of this evaluation, in finite-length analysis, we derive upper and lower bounds
of $\log N_n^\varepsilon$ while tighter bounds are preferable.
To derive the second or third order asymptotics,
we often derive upper and lower bounds, and make their asymptotic expansion
because it is quite difficult to directly derive the asymptotic expansion of  
$\log N_n^\varepsilon$.
Indeed, if upper and lower bounds match their asymptotic expansion up to
the order $\sqrt{n}$ or $\log n$,
one might consider that the upper and lower bounds are useful.
However, we cannot say that these bounds are useful if their calculation amount is very large.
As is pointed in \cite[Table 1]{Markov}\cite[Table 1]{W-H2}, for their calculation,
this kinds of upper and lower bonds require
the calculation of the cumulant distribution function of the
the independent and identical distribution of a certain distribution related to our task.
If the distribution is binary distribution,
we can easily calculate the cumulant distribution function.
Unfortunately, in the general case, its calculation is very large.

In this paper, to resolve this problem, we propose 
the concept of semi-finite length analysis as follows.
First, we derive upper and lower bounds of $\log N_n^\varepsilon$.
Then, we make their asymptotic expansion up to the constant term like
$A_1 n + A_{2,\varepsilon}\sqrt{n}+ A_3 \log n + A_4 +O(1/\sqrt{n})$.
In this case, 
the difference between the approximation $A_1 n + A_{2,\varepsilon}\sqrt{n}+ A_3 \log n + A_4$
and the true bound is guaranteed to converge to zero.
Hence, we can say that the absolute of the difference is smaller than $1$
when $n$ is sufficiently large.
Thus, 
%when we draw the graph, we can estimate the true value of the upper or lower bound.  That is, 
from these approximations of upper and lower bounds,
we can estimate the range of the true value of $\log N_n^\varepsilon$.
Since the aim is the approximate calculation of the upper and lower bounds,
their asymptotic expansions do not necessarily need to match each other.
But, if their first order coefficients do not match each other, 
the upper and/or lower bounds are so loose that they are not useful.
We call this type of analysis the {\it semi-finite length analysis} for upper or lower bounds, which can be summarized as follows.
\begin{description}
\item[R1] We can calculate the asymptotic expansion up to the constant term of the upper bound
like $A_1^+ n + A_{2,\varepsilon}^+\sqrt{n}+ A_3^+ \log n + A_4^+ +O(1/\sqrt{n})$.
\item[R2] We can calculate the same type of expansion  of the lower bound
like $A_1^- n + A_{2,\varepsilon}^-\sqrt{n}+ A_3^- \log n + A_4^- +O(1/\sqrt{n})$.
\item[R3]
 $A_1^+=A_1^-$.
\end{description}
This kind of problem has not been discussed except for the source coding without side information \cite{strassen,K-V}.

In this paper, we address this problem for secure random number generation.
To tackle this problem, 
using several useful existing results,
we derive upper and lower bounds for 
$\log N_n^\varepsilon$ of secure keys under the constraint 
that the secrecy measure is less than $\varepsilon$
when $n$ outcomes are generated according to 
the independent and identical distribution of a certain distribution.
For their asymptotic expansion,
we employ Edgeworth expansion and strong large deviation, which were derived by Bahadur and Rao \cite{BR}.
Indeed, strong large deviation was employed for information theory in 
the papers \cite{Moulin,VFKA1,VFKA2}.
While the papers \cite{Moulin,VFKA1,VFKA2} employed saddle point approximation in addition to 
strong large deviation,
in a similar way to the papers \cite{TH,IH,IKH,HO} in other topics,
we directly use the formula for strong large deviation to calculate higher order asymptotics so that we do not employ saddle point approximation.

The next target is channel coding, which has two famous finite-length bounds,
the dependent test (DT) bound \cite[Remark 15]{HN}\cite{polyanskiy:10} and the meta converse bound \cite{Nagaoka}\cite[Section 4.6]{Hbook}\cite{polyanskiy:10}.
To discuss channel coding,
using a similar derivation based on strong large deviation and 
Edgeworth expansion,
we derive semi-finite length expansion 
in the simple binary hypothesis testing 
with two frameworks, which are related to the above two types of bounds.
In fact, these two types of bounds can be used for source coding with side information 
\cite[Theorem 7]{tomamichel:12}.
Then, applying these expansions, 
we derive upper and lower bounds for this setting in the sense of semi-finite length analysis.
In the same way,
we derive the same types of upper and lower bounds for channel coding 
when we assume a symmetric condition for channel, 
the conditional additive condition defined in \cite[Section IV]{Markov}
because this assumption brings simple derivation, which enables us to get 
an asymptotic expansion up to the constant order.
Finally, we proceed to wire-tap channel \cite{Wyner,CK79,Csiszar,H-gen}.
In this model, as pointed in \cite[Section V]{hayashi:10}\cite[Section VIII]{hayashi:10b}\cite[Conclusions]{ISIT2013}, we combine 
the results of secure random number generation and channel coding.
Then, when the wire-tap channel is degraded and 
the channels to the legitimate receiver and the eavesdropper 
are conditional additive,
we derive the desired asymptotic expansion
while the paper \cite{YSP} discusses the second order asymptotics for generic wire-tap channels.

The rest of the paper is organized as follows.
In Section \ref{section:preliminaries}, we summarize notations used in this paper.
Section \ref{S3} discusses secure random number generation.
Next, Section \ref{S4} treats simple binary hypothesis testing.
Section \ref{S4-5} treats fixed-length source coding.
by sing the result of Section \ref{S4}.
Then, using the result of Section \ref{S4}, Section \ref{S5} proceeds to channel coding.
Applying the results of Sections \ref{S3} and \ref{S5},
Section \ref{S6} addresses wire-tap channel coding.
To show the asymptotic expansion given in Sections \ref{S3} and \ref{S4},
Section \ref{S7} prepares knowledge of strong large deviation and Edgeworth expansion.
Using these tools, Sections \ref{S8} and \ref{S9} show the asymptotic expansion stated in 
Sections \ref{S3} and \ref{S4}, respectively.
Section \ref{S10} gives the conclusion.

%%%%%%%%%%% Preliminaries %%%%%%%%%%%%%%%%%%%%%%
\section{Preliminaries}
\label{section:preliminaries}
To discuss higher order asymptotics,
we need several information quantities.
In this section, we prepare notations used in this paper and prepare these information quantities.

%%%% Problem Formulation %%%%%%

\subsection{Notations}
For a set ${\cal A}$, let ${\cal P}({\cal A})$ be the set of all 
probability distribution on ${\cal A}$. It is also convenient to introduce
the set $\bar{{\cal P}}({\cal A})$ of all sub-normalized non-negative functions.
We denote the set of conditional distributions on ${\cal B}$ conditioned with
${\cal A}$ by 
${\cal P}({\cal B}|{\cal A})$. 
Given a distribution $P_A \in {\cal P}({\cal A})$,
a conditional distribution $P_{B|A} \in {\cal P}({\cal B}|{\cal A})$,
we define the joint distribution $P_{B|A} \times P_A 
\in {\cal P}({\cal A}\times {\cal B}) $
as
$P_{B|A} \times P_A (b,a):= P_{B|A} (b|a) P_A (a)$.
In particular, 
$P_{B|A=a}$ is defined as $P_{B|A=a}(b):=P_{B|A} (b,a)$.
When the conditional distribution $P_{B|A=a}$
does not depend on $a\in {\cal A}$,
this notation expresses the product distribution.
That is, $P_E\times P_A$ is defined as
$P_B\times P_A(b,a):= P_B(b) P_A(a)$.
We define the distribution $P_{B|A} \cdot P_A 
\in {\cal P}({\cal B}) $ as
$P_{B|A} \times P_A (b,a):= \sum_{a \in {\cal A}}P_{B|A} (b|a) P_A (a)$.
Given $P_{AB} \in {\cal P}({\cal A} \times {\cal B})$, 
the marginal distribution $P_A $ is defined as
$P_{A}(a):= \sum_{b \in {\cal B}}P_{AB}(a,b)$.
Also, the normalized uniform distribution on ${\cal A}$
is denoted by $U_A$.
We define the distribution $P_A^n$ on ${\cal A}^n$ as
$P_A^n(a_1, \ldots, a_n):=P_A(a_1) \cdots P_A(a_n)$.
We define the conditional distribution $P_{B|A}^n$ on ${\cal B}^n$ 
conditioned with
${\cal A}$ as
$P_{B|A}^n(b_1, \ldots, b_n|a_1, \ldots, a_n):=
P_{B|A}(b_1|a_1) \cdots P_{B|A}(b_n|a_n)$.

Further,
$\mathbb{E}_P$ 
and $\mathbb{V}_P$ 
express the expectation and the variance
under the distribution $P \in \bar{{\cal P}}({\cal A})$, respectively, as follows.
\begin{align}
\mathbb{E}_P[f(A)]:= \sum_{a \in {\cal A}} P(a)f(a), \quad
\mathbb{V}_P[f(A)]:= \sum_{a \in {\cal A}} P(a)(f(a)-\mathbb{E}_P[f(A)])^2 .
\end{align}

%\subsection{Metrics for two distributions}
\subsection{Information quantity for first order asymptotics}
Given two distributions $P, Q \in \bar{{\cal P}}({\cal A})$,
the difference between two distributions are evaluated by the variational distance defined by 
\begin{eqnarray}
\label{eq:definition-of-distance}
d(P,Q) := \frac{1}{2} \sum_{a\in {\cal A}} | P(a) - Q(a)|
= \frac{1}{2} \mathbb{E}_P \Big[\Big| 1-\frac{Q}{P}\Big|\Big].
\end{eqnarray}
Also, we use the relative entropy $D(P\|Q)$ and the entropy $H(P)$
\begin{align}
D(P\|Q):= \mathbb{E}_P \Big[\log \frac{P}{Q}\Big],
\quad
H(P):= - \mathbb{E}_P \Big[\log P \Big],
\end{align}
where throughout this paper, the base of the logarithm is chosen to be $e$.
We introduce special notations for 
distributions $P$ and $Q$ on the joint system ${\cal A}$ and ${\cal B}$.
We assume that their marginal distributions on ${\cal B}$ are the same distribution $P_B$
and their conditional distributions on ${\cal A}$ condition with ${\cal B}$
are given as $P_{A|B}$ and $Q_{A|B}$.
Then, we use the notation
$D(P_{A|B} \| Q_{A|B} | P_B):= \sum_{b}P_B(b) D(P_{A|B=b}\| Q_{A|B=b} )$.
When 
a distribution $Q$ on ${\cal A}\times {\cal B}$ is given as $P_{AB}$ and
a distribution $Q$ on ${\cal A}\times {\cal B}$ 
is given as 
$Q(a,b)= R_B(b) $ by using a distribution $R_B$ on ${\cal A}$,
we write $D(P\| Q )$ as $D(P_{AB} \| R_B)$.

For $P_{AB} \in \bar{{\cal P}}({\cal A} \times {\cal B})$ and a normalized $R_B \in {\cal P}({\cal B})$,
the conditional entropy $H(P_{AB} | R_B)$ relative to $R_B$
is defined to be $-D(P_{AB} \| R_B)$.
When $P_{AB}$ is a normalized distribution and $R_B$ is the marginal distribution $P_B$, $H(P_{AB} | P_B)$ is a non-negative value.
Then, 
we define the conditional minimum entropy relative to $R_B$ \cite{renner:05d}
\begin{eqnarray}
H_{\min}(P_{AB}|R_B) := - \log \max_{(a, b) \in \rom{supp}(P_{AB})} \frac{P_{AB}(a,b)}{R_B(b)} .\Label{LLO}
\end{eqnarray}
The conditional R\'enyi entropy of order $2$ relative to $R_B$ is defined as
\begin{eqnarray}
H_2(P_{AB}|R_B) := - \log \sum_{(a, b) \in \rom{supp}(P_{AB})} \frac{P_{AB}(a,b)^2}{R_B(b)}
\ge H_{\min} (P_{AB}|R_B)
\Label{LBD}.
\end{eqnarray}

\subsection{Information quantity for higher order asymptotics}

In this paper, to get higher order expansions,
given two distributions $P, Q \in \bar{{\cal P}}({\cal A})$,
we employ the relative entropy variance $V(P\|Q)$ 
and 
$\kappa(P\|Q)$ as
\begin{align}
V(P\|Q):= &\mathbb{V}_P \Big[\log \frac{P}{Q}\Big] \\
\kappa(P\|Q):=&
\mathbb{E}_P\Big[ \Big(\frac{-\log \frac{P}{Q}
-D(P\|Q)}{\sqrt{V(P\|Q)}}\Big)^3\Big],
\end{align}
which equals the skewness, i.e., the normalized third cumulant of $-\log \frac{P}{Q}$.
To define more complicated values, 
we employ the lattice span $d_{(P\|Q)}$ of the variable $-\log \frac{P}{Q}$,
which is defined in the beginning of Section \ref{S7}.
For example, when $-\log \frac{P}{Q}$ is a continuous variable,
the lattice span $d_{(P\|Q)}$ is zero.
Then, we define the function
$v(d)$ as
\begin{align}
v(d):= 
\left\{
\begin{array}{ll}
\log \frac{d}{1-e^{-d}} & \hbox{ when } d > 0 \\
0 & \hbox{ when } d = 0 .
\end{array}
\right.
\end{align}
To describe the constant term of the asymptotic expansion,
using $v(d_{(P\|Q)})$
and 
$\Phi(x):= 
\int_{-\infty}^{x} \frac{1}{\sqrt{2\pi}}
\exp( - \frac{x^2}{2}) dx$, we define
\begin{align}
F_{1}^\varepsilon(P\|Q) :=&
\frac{\sqrt{V(P\|Q)} \kappa(P\|Q)(\Phi^{-1}(\varepsilon)^2-1)}{6 }
+ e^{v(d_{(P\|Q)})}  \\
F_{2}^\varepsilon(P\|Q) :=&
\frac{\sqrt{V(P\|Q)} \kappa(P\|Q)(\Phi^{-1}(\varepsilon)^2-1)}{6 }+
 e^{v (d_{(P\|Q)})}  \nonumber \\
& +3 \log 2-2-\log \pi  - \log V(P\|Q) - \Phi^{-1}(\varepsilon)^2 \\
F_{3}^\varepsilon(P\|Q):= &
\frac{\sqrt{V(P\|Q)} \kappa(P\|Q)(\Phi^{-1}(\varepsilon)^2-1)}{6 }
\nonumber \\
&+\frac{7}{2}\log 2-2 -\frac{1}{2}\log \pi- \frac{1}{2}\log V(P\|Q) 
- \frac{1}{2}\Phi^{-1}(\varepsilon)^2
-v(d_{(P\|Q)}).
\end{align}

\begin{align}
F_{4}^\varepsilon(P\|Q):=&
\frac{\sqrt{V(P\|Q)} \kappa(P\|Q)(\Phi^{-1}(\varepsilon)^2-1)}{6 }
+\frac{1}{2}\log (2 \pi V(P\|Q))
+\frac{1}{2}\Phi^{-1}(\varepsilon)^2
-v(d_{(P\|Q)}) \\
F_{5}^\varepsilon(P\|Q):=&
\frac{\sqrt{V(P\|Q)} \kappa(P\|Q)(\Phi^{-1}(\varepsilon)^2-1)}{6 }
-\frac{1}{2}\log V(P\|Q)
-v(d_{(P\|Q)})-1.
\end{align}

Strassen \cite{strassen} implicitly 
used the quantity $F_{4}^\varepsilon(P\|Q)$ for the non-lattice case, i.e., for the case of $d=0$ 
in the context of source coding with no side information and hypothesis testing.
Kontoyiannis and Verd\'{u} \cite{K-V} explicitly
discussed it for the non-lattice case
in the context of source coding with no side information as \cite[(36)]{K-V}.
Also, Moulin \cite{Moulin} defined
the quantity $\sqrt{V(P\|Q)} 
\frac{\kappa(P\|Q)(\Phi^{-1}(\varepsilon)^2-1)}{6 }
+\frac{1}{2}\log (2 \pi V(P\|Q))
+\frac{1}{2}\Phi^{-1}(\varepsilon)^2
$ in \cite[(2.12) and (2.14)]{Moulin}
for the general case in the context of channel coding.

Remember that we defined $D(P_{A|B} \| Q_{A|B} | P_B)$
and $D(P_{AB} \| R_B)$ in the previous subsection.
This kind of definition is also applied to the quantities
defined in this subsection.
That is, $V(P_{AB} \| R_B)$ and $F_i^\varepsilon(P_{AB} \| R_B) $ are defined in the same way as in the previous subsection.

Although the definitions in Section \ref{section:preliminaries} assume 
that the sets ${\cal A}$ and ${\cal B}$ are discrete,
these definitions are applied to the general measurable case when 
the sets ${\cal A}$ and ${\cal B}$ are measurable sets.
In this case, 
${\cal P}({\cal A})$ is the set of probability measures on ${\cal A}$,
$\bar{\cal P}({\cal A})$ is the set of non-negative measures on ${\cal A}$,
and ${\cal P}({\cal B}|{\cal A})$
is the set of conditional probability measures $\mu_{B|A=a}$
on ${\cal B}$ conditioned with $a \in {\cal A}$.
In this case, the functions $\frac{P}{Q}$ and $\frac{Q}{P}$ are defined as the Radon-Nikodym derivatives between two measures $Q$ and $P$.
The definition \eqref{LLO} is generalized as 
\begin{eqnarray}
H_{\min}(P_{AB}|R_B) := - \log 
\inf\Big\{ \mathsf{L} \Big|
\mathsf{L} \ge \frac{P_{AB}}{R_B}(a,b) \hbox{ holds almost everywhere with respect to }
P_{AB}.
\Big\}.
\end{eqnarray}
However, the uniform distribution $U_A$ is defined only when ${\cal A}$
is discrete and finite.

\section{Secure Random Number Generation}\Label{S3}
\subsection{Problem Formulation}
Let $P_{AE} \in \bar{{\cal P}}({\cal A} \times {\cal E})$ be a sub-normalized non-negative function.
For a function $f: {\cal A} \to {\cal S}$ and the key $S = f(A)$, let
\begin{eqnarray*}
P_{SE}(s,z) = \sum_{x \in f^{-1}(s)} P_{AE}(a,e).
\end{eqnarray*}
We define the security by
\begin{eqnarray*}
d(f|P_{AE}) = d(P_{SE}, U_{{S}} \times P_E).
\end{eqnarray*}
Although the quantity $d(f|P_{AE})$ has no operational meaning 
for unnormalized $P_{AE}$, it will be used to derive bounds on $d(f|P_{AE})$
for normalized $P_{AE}$. For distribution $P_{AE} \in {\cal P}({\cal A} \times {\cal E})$ and 
security parameter $\varepsilon \ge 0$, we are interested in characterizing 
\begin{eqnarray*}
\ell^\varepsilon(P_{AE}) := \sup\{ \log |{\cal S}| ~| \exists f: {\cal X} \to {\cal S} \mbox{ s.t. } d(f|P_{AE}) \le \varepsilon \}.
\end{eqnarray*} 
The inverse function is given as
\begin{align}
\Delta(m|P_{AE}):= \inf_{f} d(f|P_{AE}) .
\end{align}

\if0
We often employ a randomized hash function $F$.
Now, we impose our randomized hash function $F$ from ${\cal A}$ to ${\cal S}$ to the universal 2 condition;
\begin{align}
\mathbb{P}\{ F(a)=F(a')\} \le 
\frac{1}{|{\cal S}|}
\end{align}
for $a\neq a' \in {\cal A}$.
Then, we consider the worst case among universal 2 functions;
\begin{align}
\bar{\Delta}(m|P_{AE}):= \max_{F} \mathbb{E}_{P_F} d(F|P_{AE}) ,
\end{align}
where the distribution $P_F$ is chosen so that $F$ is a universal 2 function.
Then, we define 
\begin{eqnarray*}
\underline{\ell}^\varepsilon(P_{AE}) := 
\sup\{ m | \bar{\Delta}(m|P_{AE}) \le \varepsilon \}.
\end{eqnarray*} 
\fi

%%%% Min Entropy Bound %%%%%%%
\subsection{Single shot Analysis}\Label{LLS}
First, we employ the smooth minimum entropy framework
that was mainly introduced and developed by Renner and 
his collaborators \cite{renner:05b,renner:05d,tomamichel:09,tomamichel:10,tomamichel:phd}.
%Throughout the paper, we assume that the base of the logarithm is $2$.
%\begin{definition}
Then, we define
\begin{align*}
H_{\min}^\varepsilon(P_{AE}|R_E)&: = \max_{Q_{AE} \in {\cal B}^\varepsilon(P_{AE})} H_{\min}(Q_{AE}|R_E) ,\\
\delta_{\min}(m | P_{AE}|R_E)&:= \min_{Q_{AE} \in {{\cal P}}({\cal A} \times {\cal E})}
\{ d(Q_{AE}, P_{AE})|  H_{\min}(Q_{AE}|R_E) \le m \}
\end{align*}
where 
\begin{eqnarray*}
{\cal B}^\varepsilon(P_{AE}) := \left\{ Q_{AE} \in {\cal P}({\cal A} \times {\cal E}) : d(P_{AE}, Q_{AE}) \le \varepsilon \right\}.
\end{eqnarray*}
%\end{definition}

Then, we have a key lemma to derive a upper bound of $\ell^\varepsilon(P_{AE})$.
\begin{proposition}[Monotonicity \protect{\cite[Lemma 2]{ISIT2013}}]
\Label{lemma:monotonicity}
For any function $f:{\cal X} \to {\cal S}$,
$P_{AE} \in {\cal P}({\cal A} \times {\cal E})$, and $R_E \in {\cal P}({\cal E})$, we have
\begin{eqnarray*}
H_{\min}^\varepsilon(P_{SE}|R_E) \le H_{\min}^\varepsilon(P_{AE}|R_E).
\end{eqnarray*}
\hfill $\square$\end{proposition}
For readers' convenience, we give a proof in Appendix \ref{appendix:lemma:monotonicity}.

Using Proposition \ref{lemma:monotonicity}, we obtain the following proposition.
\begin{proposition}[\protect{ \cite[Theorem 1]{ISIT2013}}]\Label{T0}
For $P_{AE}\in {\cal P}({\cal A}\times {\cal E})$, we have
\begin{align}
 \ell^\varepsilon(P_{AE})
\le H_{\min}^{\varepsilon}(P_{AE}|P_E) \Label{LGT-1}.
\end{align}
The inequality is equivalent to 
\begin{align}
\Delta(m|P_{AE}) \ge
\delta_{\min}(m | P_{AE}|R_E).\Label{LFUT}
\end{align}
\hfill $\square$\end{proposition}

Since the paper \cite{ISIT2013} skips the detail proof of Proposition \ref{T0}, 
we give its proof for reader's convenience.
\begin{IEEEproof}[Proof of Proposition \ref{T0}]
Let $f: {\cal A}\to {\cal S} $ be a function to achieve 
the bound $\ell^\varepsilon(P_{AE})$.
Then, the resultant distribution $P_{SE}$ satisfies 
$d(P_{SE}, U_{{S}} \times P_E)=\varepsilon$.
Since $H_{\min}(U_{{S}} \times P_E|P_E)=
\log |{\cal S}|=\ell^\varepsilon(P_{AE})$,
we have $H_{\min}^\varepsilon(P_{SE}|R_E) \ge \ell^\varepsilon(P_{AE}) $.
Thus, Proposition \ref{lemma:monotonicity} yields \eqref{LFUT}.
\end{IEEEproof}

To derive the opposite evaluation, we introduce 
\begin{align}
\Delta_{\min}(m|P_{AE}|R_E)
&:= 
\min_{Q_{AE}\in \bar{{\cal P}}({\cal A} \times {\cal E})}
2 d(Q_{AE},P_{AE})
 + \frac{1}{2} \sqrt{e^{m- {H}_{\min}(Q_{AE}|R_E)}} \\
\ell_{\min}^{\varepsilon}(P_{AE}|R_E) &:= \max\{m|\Delta_{\min}(m|P_{AE}|R_E) \le \varepsilon\} .
\end{align}
To improve the evaluation, using 
the conditional R\'{e}nyi entropy of order 2,
we define 
\begin{align}
\Delta_{2}(m|P_{AE}|R_E)
&:= 
\min_{Q_{AE}\in \bar{{\cal P}}({\cal A} \times {\cal E})}
2 d(Q_{AE},P_{AE})
 + \frac{1}{2} \sqrt{ e^{m- {H}_{2}(Q_{AE}|R_E)}} \\
\ell_2^{\varepsilon}(P_{AE}|R_E) &:= \max\{m|\Delta_{2}(m|P_{AE}|R_E) \le \varepsilon\} .
\end{align}
Then, we obtain the following opposite evaluation.

\begin{proposition}[\protect{\cite[Corollary 2]{ISIT2013}\cite[Lemma 23]{hayashi:12d}\cite[Proposition 23]{epsilon}}]\Label{T1}
For $P_{AE}\in {\cal P}({\cal A}\times {\cal E})$
and $R_{E}\in {\cal P}({\cal E})$, we have
\begin{align}
\Delta_{\min}(m|P_{AE}|R_E)
\ge \Delta_{2}(m|P_{AE}|R_E)
%\ge \overline{\Delta}(m|P_{AE}) 
\ge\Delta(m|P_{AE}) .
\Label{LGT}
\end{align}
\hfill $\square$\end{proposition}
For readers' convenience, we give a proof in Appendix \ref{A2}.

\if0
Further, we define 
\begin{align}
\overline{\Delta}^*(m|P_{AE})
&:= \max_{\Omega}( P_{AE}(\Omega)- 2\cdot e^{-m} P_{E}(\Omega)) \\
\underline{\ell}^{\varepsilon,*}(P_{AE})
&:=
\sup\{ m | \overline{\Delta}^*(m|P_{AE}) \le \varepsilon \} \\
&=
\sup\{ m | \forall \Omega , 
P_{AE}(\Omega)- 2\cdot e^{-m} P_{E}(\Omega)
 \le \varepsilon \}.
\end{align}

\begin{theorem}\Label{T1-2}
For $P_{AE}\in {\cal P}({\cal A}\times {\cal E})$
and $R_{E}\in {\cal P}({\cal E})$, we have
\begin{align}
\overline{\Delta}(m|P_{AE}) 
\ge
\overline{\Delta}^*(m|P_{AE})
\Label{LGT4}
\end{align}
\hfill $\square$\end{theorem}
\fi

Combining Propositions \ref{T0}, \ref{T1}, %and \ref{T1-2}, 
we have the following evaluation.
\begin{align}
&\ell_{\min}^\varepsilon(P_{AE}|R_E)
\le \ell_2^\varepsilon(P_{AE}|R_E)
%\le \underline{\ell}^\varepsilon(P_{AE})
\le \ell^\varepsilon(P_{AE})
\le H_{\min}^{\varepsilon}(P_{AE}|P_E)\Label{LGT2} 
%\\ & \underline{\ell}^\varepsilon(P_{AE}) \le \underline{\ell}^{\varepsilon,*}(P_{AE})
\end{align}
for any distribution $R_E \in {\cal P}({\cal E})$.

\subsection{Semi-finite block-length Analysis}\Label{subsection:min-entropy}
To calculate the above upper and lower bounds, 
it is important to evaluate the values
$\frac{(P_{AE})^2}{P_E}
\Big\{ -\log \frac{P_{AE}(a,e)}{P_E(e)} >  m \Big \}$
and 
$P_{AE}
\Big\{ -\log \frac{P_{AE}(a,e)}{P_E(e)} \le  m \Big \}$ for a given value $m$.
In the asymptotic situation,
strong large deviation is known as a method to precisely evaluate 
the first quantity 
and
Edgeworth expansion is a method to evaluate the difference between
the Gaussian distribution and the second value.
Combining these two methods,
we obtain semi-finite block-length analysis
for the lower and upper bounds of $\ell^\varepsilon(P_{AE}^n)$
as follows.

\begin{theorem}\Label{T2}
For $P_{AE}\in {\cal P}({\cal A}\times {\cal E})$, we have
\begin{align}
{H}_{\min}^\varepsilon(P_{AE}^n|P_E^n)
&=nH(P_{AE}|P_E) +\sqrt{n}\sqrt{V(P_{AE}\|P_E)}\Phi^{-1}(\varepsilon)+ F_{1}^\varepsilon(P_{AE}\|P_E)+O(\frac{1}{\sqrt{n}}) 
\Label{GS1}
\\
{\ell}_{\min}^\varepsilon(P_{AE}^n|P_E^n)
&=nH(P_{AE}|P_E) +\sqrt{n}\sqrt{V(P_{AE}\|P_E)}\Phi^{-1}(\varepsilon)-\log n+ F_{2}^\varepsilon(P_{AE}\|P_E)+O(\frac{1}{\sqrt{n}}) 
\Label{GS2}
\\
{\ell}_{2}^\varepsilon(P_{AE}^n|P_E^n)
&\ge nH(P_{AE}|P_E) +\sqrt{n}\sqrt{V(P_{AE}\|P_E)}\Phi^{-1}(\varepsilon)-\frac{1}{2}\log n+ F_{3}^\varepsilon(P_{AE}\|P_E)+O(\frac{1}{\sqrt{n}}).
\Label{GS3} 
\end{align}
\if0
\underline{\ell}^{\varepsilon,*}(P_{AE}^n)
&= nH(P_{AE}|P_E) +\sqrt{n}\sqrt{V(P_{AE}\|P_E)}\Phi^{-1}(\varepsilon)-\frac{1}{2}\log n+ F_{4}^\varepsilon(P_{AE}\|P_E)+O(\frac{1}{\sqrt{n}}).
\Label{GS4} 
\fi
\hfill $\square$\end{theorem}
Theorem \ref{T2} is shown in Section \ref{S8}.
\if0
Therefore, we conclude that
\begin{align}
\underline{\ell}^{\varepsilon}(P_{AE}^n)
&= nH(P_{AE}|P_E) +\sqrt{n}\sqrt{V(P_{AE}\|P_E)}\Phi^{-1}(\varepsilon)-\frac{1}{2}\log n+ O(1).
\end{align}
\fi

\subsection{Numerical comparison}\Label{S3-6}
We numerically compare our result with the previous results \cite{ISIT2013}.
The paper \cite[Theorem 2]{ISIT2013} derived lower and upper bounds of 
$ \ell^\varepsilon(P_{AE})$ as
\begin{align}
& \max_{R_E \in {\cal P}({\cal E})}
H_{\rm sp}^{\epsilon-\eta}(P_{ZE}|R_E)+\log 4 \eta^2-1\nonumber \\
\le &
 \ell^\varepsilon(P_{AE})% \nonumber \\
\le 
{H}_{\min}^\varepsilon(P_{AE}|P_E)
\le 
H_{\rm sp}^{\epsilon-\zeta}(P_{ZE}|P_E)-\log \zeta.
\Label{W1}
\end{align}
where
\begin{align}
H_{\rm sp}^{\epsilon}(P_{ZE}|R_E):=
\inf_m \Big\{m\Big|
P_{AE}
\Big\{ -\log \frac{P_{AE}(a,e)}{P_E(e)} \le  m \Big \}
 \le \epsilon\Big \}.
\end{align}
Modifying leftover hash lemma (Proposition \ref{T1})\cite{HILL,bennett:95},
the paper \cite[Theorem 6]{ISIT2013} 
yields the following lower bound.
\begin{align}
\max_{0 \le \theta \le 1}
&\max_{R_E \in {\cal P}({\cal E})}
\theta H_{1+\theta}(P_{ZE}|R_E) \nonumber \\
&\qquad+(1+\theta) H_{\rm sp}^{\epsilon-\eta}(P_{ZE}|R_E)
+\log 2 \eta^2-1
\nonumber \\
\le &
 \ell^\varepsilon(P_{AE}).
\Label{W2}
 \end{align}

\begin{figure}[t]
    \centering
    \includegraphics[width=0.5\hsize]{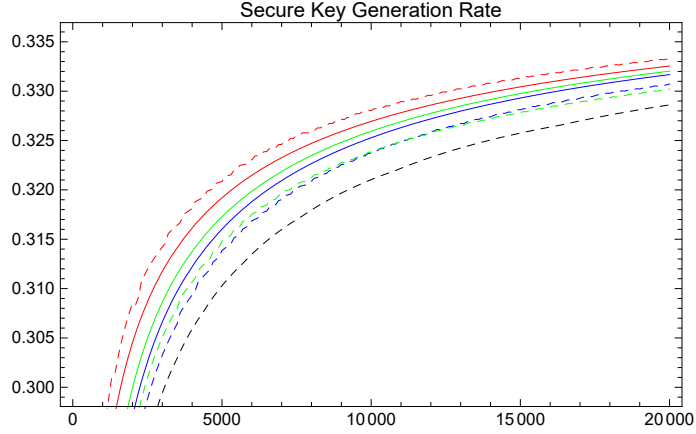}
    \caption{Upper and lower bounds when $\epsilon=0.001$ and $q=0.11$. 
    The horizontal axis expresses the block-length $n$, and the vertical axis expresses the secure key generation rate. 
    Red solid curve expresses the upper bound $\frac{1}{n} {H}_{\min}^\varepsilon(P_{AE}^n|P_E^n)
$, 
    Green solid curve expresses the lower bound $\frac{1}{n} {\ell}_{2}^\varepsilon(P_{AE}^n|P_E^n)
$, and
    Blue solid curve expresses the lower bound $\frac{1}{n} {\ell}_{\min}^\varepsilon(P_{AE}^n|P_E^n)
$.
    Red dashed curve expresses the upper bound given in \eqref{W1}, 
    Green dashed curve expresses the lower bound given in \eqref{W2},
      Blue dashed curve expresses the lower bound given in \eqref{W1},
and      Black dashed curve expresses the lower bound given in \eqref{W3}.      }
    \Label{COM}
\end{figure}

Also, using an exponential upper bound 
of leaked information $\Delta(m|P_{AE})$ given in \cite{hayashi:10},
the paper \cite[Theorem 5]{ISIT2013} derived 
the following lower bound.
\begin{align}
\sup_{0 \le \theta \le 1}
\frac{\theta H_{1+\theta}(P_{ZE}|P_E)+(1+\theta)\log 2 \epsilon/3
}{\theta}-1
%\nonumber \\
\le 
 \ell^\varepsilon(P_{AE}).
\Label{W3}
 \end{align}

Now, we consider the case such that $E$ is obtained from $A$ throughout BSC, i.e.,
${\cal A}={\cal E}=\mathbb{F}_2$ and 
\begin{align}
P_{AE}(a, a) =\frac{1 - q}{2},\quad
P_{AE}(a, a + 1) =\frac{q}{2}
\end{align}
In the following, all the information quantities 
equal those with ${\cal E}$ is a single element and $P_{AE}$ is given as
$P_{A}(0)=q$ and $P_{A}(1)=1-q$.
For this comparison, 
similar to \cite{ISIT2013},
we set $R_E$ to $P_E$ and $\eta,\zeta$ to $\epsilon/2$
in \eqref{W1}, \eqref{W2}, and \eqref{W3}.
When we choose $q=0.11$ and $\epsilon=10^{-3} $,
upper and lower bounds for the secure key generation rates
are calculated by changing $n$ in Fig. \ref{COM}.
When we choose $q=0.11$ and $n=3000$ and $100000$,
upper and lower bounds for the secure key generation rates
are calculated by changing $\epsilon$ 
in Figs. \ref{COM2} and \ref{COM3}, respectively.
These figures show that our bounds improve the bounds in the previous paper \cite{ISIT2013}.

\begin{figure}[t]
    \centering
    \includegraphics[width=0.5\hsize]{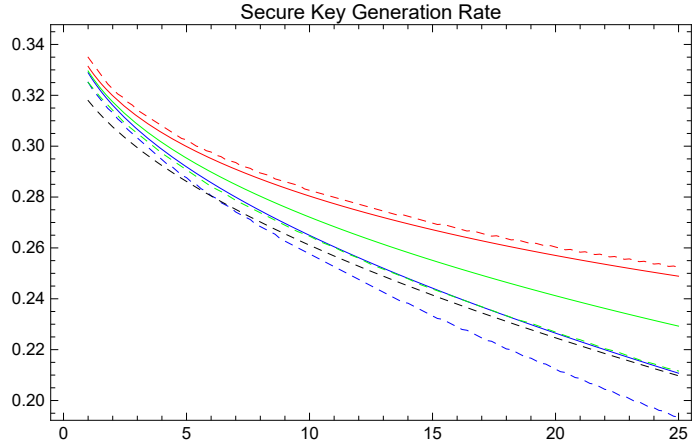}
    \caption{Upper and lower bounds when $n=3000$ and $q=0.11$. 
    The horizontal axis expresses $\log_{10}(\epsilon)$, and the vertical axis expresses the secure key generation rate. 
   Each line expresses upper or lower bound in the same way as Fig. \ref{COM}.}
    \Label{COM2}
\end{figure}

\begin{figure}[t]
    \centering
    \includegraphics[width=0.5\hsize]{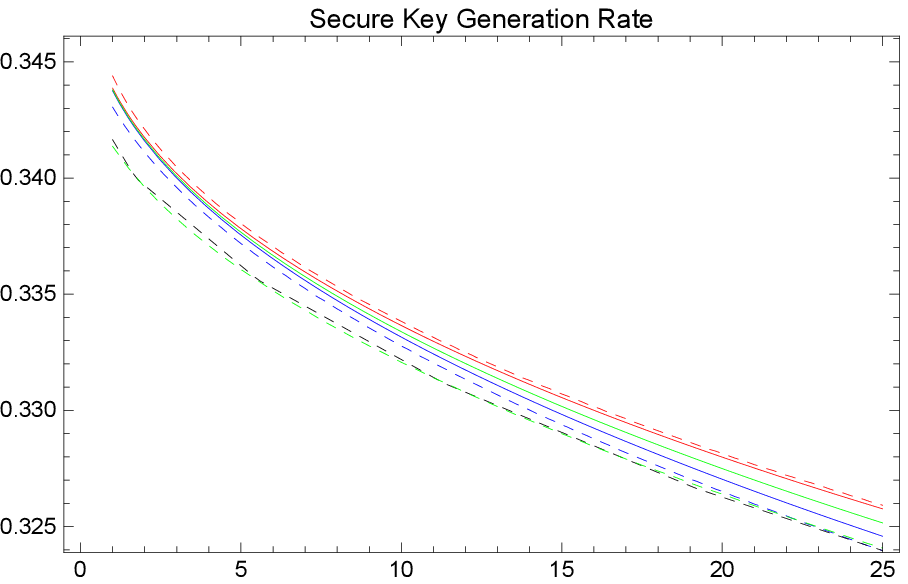}
    \caption{Upper and lower bounds when $n=100000$ and $q=0.11$. 
    The horizontal axis expresses $\log_{10}(\epsilon)$, and the vertical axis expresses the secure key generation rate. 
   Each line expresses upper or lower bound in the same way as Fig. \ref{COM}.}
    \Label{COM3}
\end{figure}

\subsection{Sacrifice bit-length}
Next, we consider sacrifice bit-length when $P_A$ is the uniform distribution.
We define the sacrifice bit-length $S^\varepsilon(A|E|P_{AE}) $ 
and its upper bounds as
\begin{align}
S^\varepsilon(A|E|P_{AE})&:=\log |{\cal A}|- \ell^\varepsilon(P_{AE})\\
S_{\min}^\varepsilon(A|E|P_{AE}|R_E)&:=\log |{\cal A}|- \ell_{\min}^\varepsilon(P_{AE}|R_E)\\
S_2^\varepsilon(A|E|P_{AE}|R_E)&:=\log |{\cal A}|- \ell_2^\varepsilon(P_{AE}|R_E) \\
I_{\max}^\varepsilon(A;E|P_{AE}|R_E)&:=\log |{\cal A}|- H_{\min}^\varepsilon(P_{AE}|R_E).
\end{align}
That is, we have
\begin{align}
S_{\min}^\varepsilon(A|E|P_{AE}|R_E)
\ge S_{2}^\varepsilon(A|E|P_{AE}|R_E)
\ge S^\varepsilon(A|E|P_{AE})
\ge I_{\max}^\varepsilon(A;E|P_{AE}|P_E).
\end{align}

Since any measure $Q_{ABE}$ satisfies 
$ H_{\min}(Q_{ABE}|R_E)=H_{\min}(Q_{ABE}|U_B\times R_E)- \log |{\cal B}|$
and
$ H_{2}(Q_{ABE}|R_E)=H_{2}(Q_{ABE}|U_B\times R_E)- \log |{\cal B}|$,
we have
\begin{align}
\ell_{2}^\varepsilon(AB|E|P_{ABE}|R_E)
&= \ell_{2}^\varepsilon(A|BE|P_{ABE}|U_B \times R_E) - \log |{\cal B}|\\
\ell_{\min}^\varepsilon(AB|E|P_{ABE}|R_E)
&= \ell_{\min}^\varepsilon(A|BE|P_{ABE}|U_B \times R_E)- \log |{\cal B}| .
\end{align}
Hence,
\begin{align}
S_{2}^\varepsilon(AB|E|P_{ABE}|R_E)
&= S_{2}^\varepsilon(A|BE|P_{ABE}|U_B \times R_E) \Label{BGJ} \\
S_{\min}^\varepsilon(AB|E|P_{ABE}|R_E)
&= S_{\min}^\varepsilon(A|BE|P_{ABE}|U_B \times R_E) .
%I_{\max}^\varepsilon(AB;E|P_{AE}|R_E)
%&= I_{\max}^\varepsilon(A;BE|P_{AE}|U_B \times R_E).
\end{align}

Therefore, we have the following lemma.
\begin{lemma}\Label{L9}
Any distribution $R_E$ of $E$ satisfies 
\begin{align}
S_{\min}^\varepsilon(AB|E|P_{ABE}|R_E)
\ge S_{2}^\varepsilon(AB|E|P_{ABE}|R_E)
\ge S^\varepsilon(A|BE|P_{ABE}).
\end{align}
\hfill $\square$\end{lemma}

\begin{IEEEproof}
The first inequality follows from the first inequality in \eqref{LGT2}.
The second inequality can be shown as follows.
\begin{align}
 S_{2}^\varepsilon(AB|E|P_{ABE}|R_E)
= S_2^\varepsilon(A|BE|P_{ABE}| U_B \times R_E)
\ge S^\varepsilon(A|BE|P_{ABE}),
\end{align}
where the first and second relations follow from 
\eqref{BGJ} and
the second inequality in \eqref{LGT2},  
respectively.

\end{IEEEproof}
\section{Hypothesis testing}\Label{S4}
Now, on a given system ${\cal A},$
we study the simple hypothesis testing problem for the null hypothesis
$P \in {\cal P}({\cal A})$
 versus the alternative hypothesis 
$Q \in {\cal P}({\cal A})$.
However, in the following,
we consider a more generalized setting for the application in the later sections.
That is, we assume that
$P \in {\cal P}({\cal A})$
and $Q \in \bar{\cal P}({\cal A})$.
In this setting, a test is given as a randomized function $T$ taking values in $[0,1]$.
When we observe $A$, 
we support $Q$ with probability $T(A)$
and does $P$ with probability $1-T(A)$.
The error probabilities of the first and the second are, respectively, defined by
\begin{align}
\alpha(T) := \mathbb{E}_{P}[1-T],\quad
\beta(T) := \mathbb{E}_{Q}[T].
\end{align}
We focus on
the minimum of the error probability of the second kind under the constant constraint of 
the error probability of the first kind;
\begin{align}
\beta_\varepsilon(P\|Q):=
\min_{T}\{ \beta(T) | \alpha(T)\le \varepsilon \}.
\end{align}
The hypothesis testing entropy is defined as
\begin{align}
D_h^\varepsilon(P\|Q):= -\log \beta_\varepsilon(P\|Q).
\end{align}

We obtain its semi-finite block-length analysis
as follows.
\begin{theorem}\Label{T3}
When $P \in {\cal P}({\cal A})$ and $Q \in \bar{\cal P}({\cal A})$, we have
\begin{align}
D_h^\varepsilon(P^n\|Q^n)
=nD(P\|Q) +\sqrt{n}\sqrt{V(P\|Q)}\Phi^{-1}(\varepsilon)
+\frac{1}{2}\log n + F_{4}^\varepsilon(P\|Q)+O(\frac{1}{\sqrt{n}}).
\Label{E27}
\end{align}
\hfill $\square$\end{theorem}
This theorem is shown in Subsection \ref{S9-A}.

\begin{remark}
In fact, Strassen \cite[Section 3]{strassen} claimed 
an asymptotic expansion for $D_h^\varepsilon(P^n\|Q^n)$.
His expansion is the same as ours in the non-lattice case.
However, his extra term caused by the lattice span $d_{(P\|Q)}$ is 
different from ours.
\if0
Theorem \ref{T3} with the non-lattice case, i.e., the case of $d=0$
is quite similar to Theorem 16 of \cite{K-V}.
The authors of \cite{K-V} mentioned that 
Strassen \cite{strassen} claimed 
a slightly stronger statement (Theorem 16 of \cite{K-V}) than 
Theorem \ref{T3} with the non-lattice case.
The authors of \cite{K-V} pointed out that
the derivation in \cite{strassen} has a gap to derive such a stronger statement.

However, the paper did not give the proof of their Theorem 16.
The paper \cite{K-V} mentioned that 
was shown in 
\fi
\end{remark}

For a preparation of the analysis of channel coding, we introduce 
the following quantity.
\begin{align}
\Delta_{DT}(m|P\|Q)&:=
\min_{\varepsilon} \varepsilon +e^{-m} e^{-D_h^\varepsilon(P\|Q)} \\
D_{DT}^\varepsilon(P\|Q)&:=
\max\{m| \Delta_{DT}(m|P\|Q) \le \varepsilon \}
\end{align}
We obtain its semi-finite block-length analysis
as follows.
%$:= E \Big[ \Big(\frac{X-E[X]}{\sqrt{V[X]}}\Big)^3\Big]$,
\begin{theorem}\Label{T4}
When $P \in {\cal P}({\cal A})$ and $Q \in \bar{\cal P}({\cal A})$, we have
\begin{align}
D_{DT}^\varepsilon(P^n\|Q^n)
=nD(P\|Q) +\sqrt{n}\sqrt{V(P\|Q)}\Phi^{-1}(\varepsilon)
+ F_{5}^\varepsilon(P\|Q)+O(\frac{1}{\sqrt{n}}).
\Label{E27B}
\end{align}
\hfill $\square$\end{theorem}

%$:= E \Big[ \Big(\frac{X-E[X]}{\sqrt{V[X]}}\Big)^3\Big]$,
This theorem is shown in Subsection \ref{S9-B}.

\begin{remark}
The paper \cite{VFKA2} made a similar analysis.
However, while they employ saddlepoint approximation in addition to strong large deviation,
our derivation is based on a simple combination of
strong large deviation and Edgeworth expansion.
\end{remark}

\section{Fixed-Length Source Coding}\Label{S4-5}
We consider fixed-length source coding.
First, we discuss the case without side information 
when the information is generated subject to the distribution $P_X$.
When we impose the decoding error probability to be not greater than $\varepsilon$,
we denote the minimum size of memory by $N^\varepsilon(X|P_X)$.
Counting the number of input elements to be correctly decoded,
we have 
\begin{align}
N^\varepsilon(X|P_X) =&
\min_{\Omega \subset {\cal X}} \{ |\Omega |  ~| P_X(\Omega) \le \varepsilon \}
=\beta_\varepsilon (P_X \| I ),
\end{align} 
where $I$ is the counting measure.
Hence, in the i.i.d. setting, 
using the asymptotic expansion given in Theorem \ref{T3} for 
$-D_h^\varepsilon(P_X^n\| I^n)$,
we have 
%the optimal coding length $\log N^\varepsilon(X^n|P_X^n)$
%is given as $-D_h^\varepsilon(P_X^n\| I^n)$, whose asymptotic expansion is given in Theorem \ref{T3}.

\begin{align}
\log N^\varepsilon(X^n|P_X^n)
= n H(P_X) -\sqrt{n}\sqrt{V(P_X\|I)}\Phi^{-1}(\varepsilon)
-\frac{1}{2}\log n 
- F_{4}^\varepsilon(P_X\|I)+O(\frac{1}{\sqrt{n}}).
\Label{KKF}
\end{align}

\if0
\begin{remark}
Strassen \cite[Theorem 1.1]{strassen} gave the same expansion as \eqref{KKF}
in the non-lattice case.
While he gave the expansion for $\log N^\varepsilon(X^n|P_X^n)$
 in the lattice case,
his expansion is different from \eqref{KKF}.
\end{remark}
\fi

\begin{remark}
Here, we should remark the relation between 
fixed-length source coding
and lossless variable-length source coding 
In lossless variable-length source coding,
we focus on the overflow probability for the respective coding length.
As pointed in \cite[Section I-D]{K-V}, 
the minium overflow probability for the respective coding length
equals $N^\varepsilon(X|P_X)$.
While the papers \cite{Kontoyiannis,K-V}
gave an asymptotic evaluation for 
the minium overflow probability for the respective coding length,
they can be regarded as the result for $N^\varepsilon(X|P_X)$.

In addition, the paper \cite[Theorems 17 and 18]{K-V}
derived similar evaluation as \eqref{KKF}.
Their evaluation is different from \eqref{KKF} in the following point.
The equation \eqref{KKF} is the asymptotic expansion, in which
we did not exactly give an upper bound of the error term.
It shows that the error term has the behavior with order $O(\frac{1}{\sqrt{n}})$.
In contrast, Theorems 17 and 18 in \cite{K-V} gave
upper and lower bounds without error.
Hence, their terms of the constant order 
 are different from \eqref{KKF}.
\end{remark}

Next, we discuss the compression of the variable $X$ with side information $Y$
when the information is generated subject to the distribution $P_{X,Y}$.
When we impose the decoding error probability to be not greater than $\varepsilon$,
we denote the minimum size of memory by $N^\varepsilon(X|Y|P_{X,Y})$.
\begin{proposition}[\protect{\cite[Theorem 7]{tomamichel:12}}]\Label{P8}
For a probability distribution $P_{X,Y} \in {\cal P}({\cal X} \times {\cal Y})$, we have 
\begin{align}
-D_h^\varepsilon(P_{X,Y}\| P_Y)
\le
\log N^\varepsilon(X|Y|P_{X,Y}) \le
-D_{DT}^\varepsilon(P_{X,Y}\| P_Y)
.\Label{HUT}
\end{align}
\hfill $\square$\end{proposition}

The first inequality of \eqref{HUT} is the same as the first inequality of 
\cite[Theorem 7]{tomamichel:12}.
The second inequality of \eqref{HUT} was essentially derived 
in the proof of the second inequality of \cite[Theorem 7]{tomamichel:12}.
For the readers' convenience, we show the second inequality of \eqref{HUT}
in Appendix \ref{A4}. 

Applying Theorems \ref{T3} and \ref{T4} to 
$-D_h^\varepsilon(P_{X,Y}^n\| P_Y^n)$
and
$-D_{DT}^\varepsilon(P_{X,Y}^n\| P_Y^n)$, respectively,
we obtain the following theorems.

\begin{theorem}\Label{T6}
For a probability distribution $P_{X,Y} \in {\cal P}({\cal X} \times {\cal Y})$, we have 
\begin{align}
& 
nH (P_{X,Y}| P_Y) -\sqrt{n}\sqrt{V(P_{X,Y}\| P_Y)}\Phi^{-1}(\varepsilon)
-\frac{1}{2}\log n - F_{4}^\varepsilon(P_{X,Y}\| P_Y)+O(\frac{1}{\sqrt{n}})
\nonumber \\
\le &
\log N^\varepsilon(X^n|Y^n|P_{X,Y}^n)
 \le
n H(P_{X,Y}| P_Y) -\sqrt{n}\sqrt{V(P_{X,Y}\| P_Y)}\Phi^{-1}(\varepsilon)
- F_{5}^\varepsilon (P_{X,Y}\| P_Y)+O(\frac{1}{\sqrt{n}}) .
\end{align}
\hfill $\square$\end{theorem}

\section{Channel Coding}\Label{S5}
\subsection{General case}
Now, we consider a channel from the input discrete system ${\cal X}$
to the output system ${\cal Y}$.
The channel is written as conditional distribution $W$.
When the input distribution is $P_X$, we denote 
the joint distribution over ${\cal X}\times {\cal Y}$ by $W\times P_X$,
and the output distribution over ${\cal Y}$ by $W\cdot P_X$.
Then, the mutual information is written as $I(X;Y)_{P_X}:=D(W\times P_X \| (W\cdot P_X)\times P_X) $.
The channel capacity is given by $C_W:=\max_{P_X} I(X;Y)_{P_X}$.
Then, we define 
\begin{align}
V_{+}:= \max_{P_X \in {\cal C}} V(X;Y)_{P_X}, \quad
V_-:=\min_{P_X \in {\cal C}} V(X;Y)_{P_X},
\end{align}
where ${\cal C}:= \{P| C_W=I(X;Y)_{P}\}$
and $ V(X;Y)_{P_X}:= D(W\times P_X \| (W\cdot P_X)\times P_X)$.
When $\varepsilon <\frac{1}{2}$, 
we define $V_\varepsilon$ to be $V_-$.
Otherwise, we define $V_\varepsilon$ to be $V_+$.
We denote the distribution in ${\cal C}$ to attain $V_\varepsilon$
by  $P_{\varepsilon}$.

Under the channel $W$,
we denote the maximum size of transmitted information with decoding error probability $\varepsilon>0$
by $N^{\varepsilon}(W)$.
Hence, when we use the channel $W$ $n$ times,
this maximum number is written as
$N^{\varepsilon}(W^n)$.
%Here, we restrict the size of message to be a power of $2$.

\begin{proposition}[\protect{\cite[Remark 15]{HN}\cite{polyanskiy:10}}]\Label{P3}
For a channel $W \in {\cal P}({\cal Y}|{\cal X})$ and a distribution 
$P_{X}\in {\cal P}({\cal X})$, 
we have 
\begin{align}
\log N^{\varepsilon}(W) \ge 
\log N^{\varepsilon}(W,P_X):= 
D_{DT}^\varepsilon ( W \times P_{X} \| (W\cdot P_{X})\times P_{X}).
\end{align}
\hfill $\square$\end{proposition}

\begin{proposition}[\protect{\cite{Nagaoka}\cite[Section 4.6]{Hbook}\cite{polyanskiy:10}}]\Label{P4}
For a channel $W\in {\cal P}({\cal Y}|{\cal X})$, we have
\begin{align}
\log N^{\varepsilon}(W) \le 
\min_{Q_{Y}\in {\cal P}({\cal Y})} \max_{P_{X}\in {\cal P}({\cal X})}
D_{h}^\varepsilon ( W \times P_{X} \| Q_{Y}\times P_{X}).
\end{align}
\hfill $\square$\end{proposition}

Substituting $P_\varepsilon^n$ and $W^n$
into $P_{X}$ and $W$ in 
Proposition \ref{P3} and Theorem \ref{T4}, 
we obtain the following theorem.
\begin{theorem}\Label{T7}
For a channel $W\in {\cal P}({\cal Y}|{\cal X})$, we have
\begin{align}
\log N^{\varepsilon}(W^n) 
\ge&  
n C_W +\sqrt{n}\sqrt{V_\varepsilon}\Phi^{-1}(\varepsilon)
 + F_{5}^\varepsilon(W\times P_\varepsilon \| (W\cdot P_\varepsilon)\times P_\varepsilon)+O(\frac{1}{\sqrt{n}}) \Label{LPY}.
\end{align}
\hfill $\square$\end{theorem}

In fact, the random coding union (RCU) bound achieves the lower bound
$n C_W +\sqrt{n}\sqrt{V_\varepsilon}\Phi^{-1}(\varepsilon)
+\frac{1}{2}\log n+ O(1)$ under some restrictions for the channel 
\cite[Corollary 54]{Polyanskiy}.
Our bound \eqref{LPY} has the following advantage over the evaluation in \cite[Corollary 54]{Polyanskiy}.
Unfortunately, their method did not identified the constant-order term.
When we need to use a lower bound whose constant-order term is determined,
we need to use \eqref{LPY} instead of the bound from the RCU bound.
given in \cite[Corollary 54]{Polyanskiy}.

\subsection{Conditional additive channel}
\subsubsection{Direct part}
On the other hand,
a channel $W$ is called conditional additive
when ${\cal Y}$ is written as ${\cal X} \times \tilde{\cal Y}$,
${\cal X}$ is an additive group, and 
the $W(x,\tilde{y}|x')=W(x-x',\tilde{y}|0)$ relation holds.
We denote the set of conditional additive channels from ${\cal X}$
to ${\cal Y}$ by
${\cal P}_a({\cal Y}|{\cal X}) $. 
For example, as shown in \cite[Section IV-C]{Markov}, 
additive Gaussian channel with the BPSK scheme 
can be regarded as a conditional additive channel.
In the following discussion, we use the notation $P_{X,\tilde{Y}}(x,\tilde{y}):=W(x,\tilde{y}|0)$.
In this case, 
an encoder is called algebraic when 
the message set is an additive group and the encoder is a homomorphism.
In $n$ uses of the channel $W$,
under the above restriction, 
we denote the maximum size of transmitted information with decoding error probability $\varepsilon>0$
by $N^\varepsilon_a(W)$.
Then, we have 
the relation $ N^\varepsilon(W) \ge  N^\varepsilon_a(W) $ and
the following proposition.
\begin{proposition}[\protect{\cite{VH},\cite[Lemma 20]{Markov}}]\Label{P3B} 
For a conditional additive channel $W \in {\cal P}_a({\cal Y}|{\cal X})$, 
the uniform distribution $U_{X}$ on ${\cal X}$ satisfies 
\begin{align}
%\log N_{\varepsilon,n}(W) \ge 
\log N^\varepsilon_a(W) \ge 
\log N^\varepsilon (W,U_X).
\end{align}
\hfill $\square$\end{proposition}
Now, we define 
$P_{Y}(y):= \sum_{x' \in {\cal X}}\frac{1}{|{\cal X}|}
W(y|x')= U_X(x) P_{\tilde{Y}}(\tilde{y})$ with $y=(x,\tilde{y})$.
Then, we apply the above discussion to the channel $W^n
\in {\cal P}({\cal Y}^n|{\cal X}^n)$.
For any $x^n \in {\cal X}^n$, we have
$D_{DT}^\varepsilon ( W_{x^n}^n  \| P_{Y}^n)
=D_{DT}^\varepsilon ( P_{X,\tilde{Y}}^n  \| P_{Y}^n)$,
where $W_{x}(y):=W(y|x) $.
Hence, 
\begin{align}
D_{DT}^\varepsilon ( W^n \times U_X^n \| (W^n\cdot U_X^n)\times U_X^n)
=D_{DT}^\varepsilon ( W_{0}^n  \| W^n\cdot U_X^n)
=D_{DT}^\varepsilon ( P_{X,\tilde{Y}}^n  \| P_{Y}^n).
\Label{HER}
\end{align}
Using Proposition \ref{P3B}, the first equation in \eqref{HER}, and Theorem \ref{T4},
we have the following theorem.
\begin{theorem}\Label{T8}
For a conditional additive channel $W\in {\cal P}_a({\cal Y}|{\cal X})$, we have
\begin{align}
& \log N^\varepsilon_a(W^n) 
\ge D_{DT}^\varepsilon ( W_{0}^n  \| W^n\cdot U_X^n) \nonumber \\
=& n C(W)
+\sqrt{n}\sqrt{V(W_{0}  \| W\cdot U_X}\Phi^{-1}(\varepsilon)
 + F_{5}^\varepsilon (W_{0}  \| W\cdot U_X)+O(\frac{1}{\sqrt{n}}) .
\Label{MBT}
\end{align}
\hfill $\square$\end{theorem}

To get another expression of \eqref{MBT} based on the structure 
${\cal Y}={\cal X}\times \tilde{\cal Y},{\cal Z}={\cal X}\times \tilde{\cal Z}$.
we see
$D( P_{X,\tilde{Y}} \| P_{Y})=D(P_{X|\tilde{Y}}\| U_X| P_{\tilde{Y}} )
= \log |{\cal X}| - H(  P_{X,\tilde{Y}}|P_{\tilde{Y}} )$.
Similarly, we have 
$V( P_{X,\tilde{Y}} \| P_{Y})= V(  P_{X,\tilde{Y}} \|P_{\tilde{Y}} )$,
$\kappa( P_{X,\tilde{Y}} \| P_{Y})= \kappa(  P_{X,\tilde{Y}} \|P_{\tilde{Y}} )$,
and
$F_{i}^\varepsilon( P_{X,\tilde{Y}} \| P_{Y})= F_{i}^\varepsilon(  P_{X,\tilde{Y}} \|P_{\tilde{Y}} )$ for $i=1,\ldots, 5$.
Using Proposition \ref{P3B}, the second equation in \eqref{HER}, Theorem \ref{T4}, and these relations,
we have the following theorem.
\begin{theorem}\Label{T9}
For a conditional additive channel $W\in {\cal P}_a({\cal Y}|{\cal X})$, we have
\begin{align}
& \log N^\varepsilon_a(W^n) 
\ge D_{DT}^\varepsilon ( P_{X,\tilde{Y}}^n  \| P_{Y}^n) \nonumber \\
=& n (\log |{\cal X}| - H(  P_{X,\tilde{Y}}|P_{\tilde{Y}} ))
+\sqrt{n}\sqrt{V(P_{X,\tilde{Y}}\|P_{\tilde{Y}})}\Phi^{-1}(\varepsilon)
 + F_{5}^\varepsilon(P_{X,\tilde{Y}}\|P_{\tilde{Y}})+O(\frac{1}{\sqrt{n}}) .
\end{align}
\hfill $\square$\end{theorem}

\subsubsection{Converse part}
Similarly,
for any $x^n \in {\cal X}^n$, we have
$D_{h}^\varepsilon ( W_{x}^n  \| P_{Y}^n)
=D_{h}^\varepsilon ( P_{X,\tilde{Y}}^n  \| P_{Y}^n)$.
Hence, 
$
D_{h}^\varepsilon ( W^n \times U_X^n \| (W^n\cdot U_X^n)\times U_X^n)
=
D_{h}^\varepsilon ( P_{X,\tilde{Y}}^n  \| P_{Y}^n)$.
Choosing $Q_{Y^n}  $ to be $P_{Y}^n $, we have
\begin{align}
D_{h}^\varepsilon ( W^n \times P_{X^n} \| P_{Y}^n\times P_{X^n})
=
D_{h}^\varepsilon ( P_{X,\tilde{Y}}^n  \| P_{Y}^n).\Label{NFT}
\end{align}
Using Proposition \ref{P4}, \eqref{NFT}, and Theorem \ref{T3}, 
we have the following theorem.
\begin{theorem}\Label{T10}
For a conditional additive channel $W\in {\cal P}_a({\cal Y}|{\cal X})$, we have
\begin{align}
& \log N^\varepsilon_a(W^n) 
\le \log N^\varepsilon(W^n) \le 
D_{h}^\varepsilon ( P_{X,\tilde{Y}}^n  \| P_{Y}^n)\nonumber \\
=& n (\log |{\cal X}| - H(  P_{X,\tilde{Y}}|P_{\tilde{Y}} ))
+\sqrt{n}\sqrt{V(P_{X,\tilde{Y}}\|P_{\tilde{Y}})}\Phi^{-1}(\varepsilon)
+\frac{1}{2}\log n
 + F_{4}^\varepsilon(P_{X,\tilde{Y}}\|P_{\tilde{Y}})+O(\frac{1}{\sqrt{n}}) .
\end{align}
\hfill $\square$\end{theorem}

The paper \cite[Theorem 55]{Polyanskiy} derived the same evaluation up to the order $\log n$
under the weakly input-symmetric condition \cite[Definition 9]{Polyanskiy}, which is similar to the conditional additive condition, but is different from it.
The same relation was shown for general DMS channels in \cite[Theorem 1]{TT1} 
and for AWGN channel with energy constraint in \cite[Theorem 65]{polyanskiy:10}.
Their achievability was shown in \cite[Corollary 54]{Polyanskiy} and \cite[(9)]{TT2},
respectively.
However, they did not derive the constant term of the upper bound.

The combination of the results \cite{Polyanskiy} and \cite{TT1}
gives the tight evaluation up to the order $\log n$.
However, their method does not give the evaluation of the error of the order $O(1)$.
Since the error of the order $O(1)$ cannot be estimated,
their method cannot give the possible range of the maximum transmittable length $\log N^\varepsilon_a(W^n)$
precisely.
On the other hand, the combination of Theorems \ref{T8} (\ref{T9}) and \ref{T10}
gives upper and lower bounds for the maximum transmittable length $\log N^\varepsilon_a(W^n)$
with small error $o(1/\sqrt{n})$.
 Since the error of the order $o(1/\sqrt{n})$ is guaranteed to converge to zero,
 our method gives the possible range of the true value of 
 $\log N^\varepsilon_a(W^n)$
 when $n$ is sufficiently large.

\if0
\section{Channel Resolvability}\Label{A3-5}
We discuss the channel resolvability for 
conditional additive channels.
\fi

\section{Wire-tap Channel Coding}\Label{S6}
\subsection{Direct part}
Next, we consider wire-tap channel coding,
in which, there are three players, 
the sender, and the legitimate receiver, and the eavesdropper.
The input system of the sender is written by ${\cal X}$,
and
the systems of the legitimate receiver and the eavesdropper are written by ${\cal Y}$ and ${\cal Z}$,
respectively.
The channels to the legitimate receiver and the eavesdropper are
are written as conditional distributions $W_Y$ and $W_Z$ on ${\cal Y}$ and ${\cal Z}$
with conditioned to ${\cal X}$, respectively.
Hence, a pair of channels $(W_Y,W_Z)$ gives a wire-tap channel.
In $n$ uses of these channels, 
the quality of this task for a wire-tap code $\phi$ is characterized by the decoding error probability 
$\varepsilon(\phi)$
and the following secrecy measure
\begin{align}
\delta(\phi) :=d(  P_{M,Z } , P_M \times P_{Z} ),
\end{align}
where $P_{M,Z }$ is the joint distribution of the message $M$ and 
the eavesdropper's information $Z$,
and $P_M \times P_{Z} $ is the product distribution of the marginal distributions
$P_M$ and $P_{Z} $.
%In $n$ uses of these channels,
%we denote the maximum size of transmitted information under the conditions $ \varepsilon(\phi)\le \varepsilon$ and $\delta(\phi)\le \delta $
%by $N_{\varepsilon,\delta,n}(W_Y,W_Z)$.

First, we discuss the performance when a wire-tap code is constructed from 
a specific algebraic error correcting code.
We assume that 
${\cal X}$ is an additive group whose order is a prime power
 and a wire-tap channel $W_Y,W_Z$ is conditional additive.
Then, ${\cal Y}$ and ${\cal Z}$ are written as 
${\cal X}\times \tilde{\cal Y}$
and
${\cal X}\times \tilde{\cal Z}$, respectively.
In the following, we employ two distribution
$P_{X,\tilde{Y}}$ and $P_{X,\tilde{Z}}$ defined as
$P_{X,\tilde{Y}}(x,\tilde{y}):=W_Y(x,\tilde{y}|0)$
and
$P_{X,\tilde{Z}}(x,\tilde{z}):=W_Z(x,\tilde{z}|0)$.
Then, we impose the algebraic condition to our code.
In $n$ uses of these channels, under this restriction for codes,
we choose an algebraic error correcting code 
with an encoder $\phi_e$ and a decoder $\phi_d$,
and denote its coding size $N_c{(\phi_e,\phi_d)}$.
Then, we denote the message of this code by $\tilde{M}$
and the set of messages by $\tilde{{\cal M}} $.
When the sender sends the message $\tilde{M}$,
we denote the joint distribution of  $\tilde{M} $ and the eavesdropper's information $Z^n$
by $P_{\tilde{M},Z^n}$.
Since the size of the set ${\cal X}$ is a prime power, 
that of ${\cal M}$ is also a prime power.
Hence, we can choose sets ${\cal M} $ and ${\cal L}$ such that
$\log |{\cal L}|={\ell^\delta(P_{\tilde{M},Z^n})}$ and 
$|{\cal M}|\cdot |{\cal L}|=\tilde{{\cal M}} $.
First, we choose a hash function $f$ from $\tilde{{\cal M}}\to {{\cal M}} $
such that $d(f| P_{\tilde{M},Z^n})\le \delta$.
We choose a function $g_f$ from ${\cal M}\cdot {\cal L}$ to 
$\tilde{{\cal M}}$ such that $f(g_f(m,l))=m$.
Then, we define the encoder and the decoder for wire-tap channel as follows.
When the sender intended to transmit the message $m$,
she generates the random variable $L$ subject to the uniform distribution on ${\cal L}$,
and transmit $ \phi_e(g_f(m,L)).$  
The legitimate receiver apply the function $f\circ \phi_d $ to the received information $Y^n$.
We denote this code by $\phi(f,\phi_e,\phi_d)$
As explained in \cite[Appendix A]{VCH}, this kind of code construction is practical
because we can choose an error correcting code with small decoding complexity and the implementation of 
$g_f$ is also easy.
Then, we have
\begin{align}
 \varepsilon(\phi(f,\phi_e,\phi_d))\le \varepsilon
\hbox{ and } \delta(\phi(f,\phi_e,\phi_d))\le \delta .
\Label{MFU}
\end{align}
Denoting the size of the message of the wire-tap code $\phi$ by $N_w(\phi)$, we have
\begin{align}
& \log N_w(\phi(f,\phi_e,\phi_d))
=\log N_c{(\phi_e,\phi_d)}- S^\delta(\tilde{M}|Z^n |P_{\tilde{M},Z^n}).\Label{GOU}
\end{align}

Now, we consider the situation when the sender generates $X^n$ subject to $U_X^n$.
Then, we identify $\tilde{\cal M}$ with an additive subgroup of ${\cal X}^n$.
We can define the quotient set ${\cal X}^n/ \tilde{\cal M}$.
For any element $X^n \in {\cal X}^n$, we denote the coset in ${\cal X}^n/ \tilde{\cal M}$
which contains $X^n$ by $[X^n]$.
When the sender inform the coset inform the information $[X^n]$ 
to the eavesdropper,
due to the algebraic structure, the information leakage does not depends on the value $[X^n]$.
Hence, we have 
\begin{align}
S^\delta(\tilde{M}|Z^n |P_{\tilde{M},Z^n})
= S^\delta(X^n|[X^n], Z^n | W_Z^n \times U_X^n) 
\le S_2^\delta(X^n|Z^n | W_Z^n \times U_X^n|R_Z^n),\Label{LGY}
\end{align}
where the inequality follows from Lemma \ref{L9}.
We denote ${\cal Z}$ by ${\cal X}' \times \tilde{\cal Z}$,
where we use the notation ${\cal X}'$ to 
distinguish the set from the input system ${\cal X}$
while it is the same set as ${\cal X}$.
Thus, we have
\begin{align}
&\log N_w(\phi(f,\phi_e,\phi_d))
\stackrel{(a)}{\ge} 
\log N_c{(\phi_e,\phi_d)}- 
 S_2^\delta(X^n|Z^n | W_Z^n \times U_X^n|(U_{X'} \times P_{\tilde{Z}} )^n)
\nonumber \\
\stackrel{(b)}{=}&
\log N_c{(\phi_e,\phi_d)}- 
n C(W_Z)
 +\sqrt{n}
 ( \sqrt{V(W_{Z|X=0}\| W_Z\cdot U_X)}\Phi^{-1}(\delta)) \nonumber \\
& -\frac{1}{2}\log n
 + F_{3}^\delta (W_{Z|X=0}\| W_Z\cdot U_X)
  +O(\frac{1}{\sqrt{n}}) ,\Label{LMY4}
\end{align}
where
the inequality $(a)$ follows from \eqref{GOU} and \eqref{LGY},
and the equation $(b)$ follows from 
the combination of 
\eqref{LGT2} and \eqref{GS3} in Theorem \ref{T2} with substituting $ U_{X'} \times P_{\tilde{Z}}=  W_Z\cdot U_X$ into $R_{Z}$.
In the derivation, we also use the relations
$V(W_Z \times U_X\|U_{X'} \times P_{\tilde{Z}})
=V(W_{Z|X=0}\| W_Z\cdot U_X) $
and
$F_{3}^\delta(W_Z \times U_X\|U_{X'} \times P_{\tilde{Z}})
=F_{3}^\delta(W_{Z|X=0}\| W_Z\cdot U_X) $.

We rewrite the above evaluation by using the structure 
${\cal Y}={\cal X}\times \tilde{\cal Y},{\cal Z}={\cal X}\times \tilde{\cal Z}$.
When $R_Z= U_{X'} \times P_{\tilde{Z}}  $, 
\begin{align}
 S_2^\delta(X^n|Z^n | W_Z^n \times U_X^n|(U_{X'} \times P_{\tilde{Z}} )^n)
 =
 S_2^\delta(X^n| {X'}^n,\tilde{Z}^n |
  W_Z^n \times U_X^n|U_{X'}^n \times P_{\tilde{Z}^n})
=
 S_2^\delta(X^n| \tilde{Z}^n |
P_{X,\tilde{Z}}^n| P_{\tilde{Z}}^n),\Label{LGY2}
\end{align}
where the second equation follows from the relation
$W_Z^n \times U_X^n(x,x',\tilde{z})=
\frac{1}{|{\cal X}|}P_{X,\tilde{Z}}(x'-x,\tilde{z})$.
Therefore, we have 
\begin{align}
&\log N_w(\phi(f,\phi_e,\phi_d))
\stackrel{(a)}{\ge} 
\log N_c{(\phi_e,\phi_d)}- 
 S_2^\delta(X^n| \tilde{Z}^n |
P_{X,\tilde{Z}}^n| P_{\tilde{Z}}^n)
\nonumber \\
\stackrel{(b)}{=}  &
\log N_c{(\phi_e,\phi_d)}- 
n (\log |{\cal X}|- H(P_{X,\tilde{Z}}| P_{\tilde{Z}}) )
 +\sqrt{n}
 \sqrt{V(P_{X,\tilde{Z}}\| P_{\tilde{Z}})}\Phi^{-1}(\delta) \nonumber \\
& -\frac{1}{2}\log n
+ F_{3}^\delta (P_{X,\tilde{Z}}\| P_{\tilde{Z}})
  +O(\frac{1}{\sqrt{n}}) ,\Label{LMY5}
\end{align}
where
the inequality $(a)$ follows from \eqref{GOU}, \eqref{LGY}, and \eqref{LGY2},
and the equation $(b)$ can be shown in the same way as \eqref{LMY4}.

Now, we proceed the analysis on the optimal performance.
For this aim, we denote the maximum size 
of transmitted information under the conditions
$ \varepsilon(\phi)\le \varepsilon$ and $\delta(\phi)\le \delta $
by $N^{\varepsilon,\delta}(W_Y,W_Z)$. 
The aim of the following discussion is to evaluate $N^{\varepsilon,\delta}(W_Y^n,W_Z^n)$.
We prepare an algebraic code for channel coding with $n$ use of the channel $W_Y$
with the decoding error probability $\varepsilon$ and the size of message 
$N^\varepsilon_a(W_{Y}^n,U_X^n)$.
We apply this algebraic code to the above wire-tap code construction, which satisfies the condition \eqref{MFU}.
Using \eqref{LMY4} and Theorem \ref{T8},
we have the following theorem.

\begin{theorem}\Label{T11}
When ${\cal X}$ is an additive group whose order is a prime power
 and a wire-tap channel $W_Y,W_Z$ is conditional additive,
\begin{align}
&\log N^{\varepsilon,\delta}(W_Y^n,W_Z^n)
\nonumber \\
{\ge} 
&
n (C(W_Y)-C(W_Z))
 +\sqrt{n}
 (\sqrt{V(W_{Y|X=0} \| W_Y\cdot U_X)}
 \Phi^{-1}(\varepsilon)+
 \sqrt{V(W_{Z|X=0}\| W_Z\cdot U_X)}\Phi^{-1}(\delta)) \nonumber \\
& -\frac{1}{2}\log n
 + F_{5}^\varepsilon(W_{Y|X=0} \| W_Y\cdot U_X)
+ F_{3}^\delta (W_{Z|X=0}\| W_Z\cdot U_X)
  +O(\frac{1}{\sqrt{n}}) .\Label{LMY0}
\end{align}
\hfill $\square$\end{theorem}
Using \eqref{LMY4} and Theorem \ref{T9},
we have the following theorem.
\begin{theorem}\Label{T12}
Under the same assumption as Theorem \ref{T11}, 
we have
\begin{align}
&\log N^{\varepsilon,\delta}(W_Y^n,W_Z^n)
\nonumber \\
\ge  &
n (H(P_{X,\tilde{Z}}| P_{\tilde{Z}})
-H(P_{X,\tilde{Y}}| P_{\tilde{Y}}))
 +\sqrt{n}
 (\sqrt{V(P_{X,\tilde{Y}}\| P_{\tilde{Y}})}\Phi^{-1}(\varepsilon)+
 \sqrt{V(P_{X,\tilde{Z}}\| P_{\tilde{Z}})}\Phi^{-1}(\delta)) \nonumber \\
& -\frac{1}{2}\log n
 + F_{5}^\varepsilon(P_{X,\tilde{Y}}\| P_{\tilde{Y}})
+ F_{3}^\delta (P_{X,\tilde{Z}}\| P_{\tilde{Z}})
  +O(\frac{1}{\sqrt{n}}) .\Label{LMY}
\end{align}
\hfill $\square$\end{theorem}

The paper \cite[Theorem 13]{YSP} derived a similar evaluation as \eqref{LMY0} and \eqref{LMY} up to the order of $\sqrt{n}$
in generic channels.
While the discussed more generic channels, they did not derived the constant term.

\subsection{Converse part}
A wire-tap channel $(W_Y,W_Z)$ is called degraded when
there exists conditional distribution $W_{Z|Y}$ on ${\cal Z}$ with conditioned to ${\cal Y}$ such that 
$W_Z(z|x)=  \sum_{y \in {\cal Y}}W_{Z|Y}(y|z)W_Y(y|x)$.
In this case, 
we define the joint condition distribution 
$\tilde{W}_{YZ}(y,z|x):=W_{Z|Y}(z|y)W_Y(y|x)$.
Indeed, there is a possibility that the true joint condition distribution 
$W_{YZ}$ is different from $\tilde{W}_{YZ}$.
Since our metric $\varepsilon(\phi)$ and 
$\delta(\phi)$ depend on $W_Y$ and $W_Z$,
we do not need to care about the form of 
the true joint condition distribution $W_{YZ}$.

\begin{proposition}[\protect{\cite[Theorem 6]{Allerton2014}}]\Label{P4-B}
When a wire-tap channel $W_Y,W_Z$ is degraded,
we have
\begin{align}
\log N^{\varepsilon,\delta}(W_Y,W_Z)
\le \min_{Q_{Y|Z}\in {\cal P}({\cal Y}|{\cal Z}),
 Q_{Z|X}\in {\cal P}({\cal Z}|{\cal X})} 
 \max_{P_{X}\in {\cal P}({\cal X})}
D_{h}^{\varepsilon+\delta} ( \tilde{W}_{YZ}\times P_{X} \| 
Q_{Y|Z}\times Q_{Z|X}\times P_{X}).
\end{align}
\hfill $\square$\end{proposition}

\begin{remark}
Originally, the paper \cite{Allerton2014} showed 
the above statement as their Theorem 6 when feedback is allowed.
Since the distribution $P_{X}$ corresponds to our code,
the case with the distribution $P_{X}$ with no feedback
corresponds to the case when no feedback is allowed.
Hence, Theorem 6 of \cite{Allerton2014} yields the above statement.
\end{remark}

We define $ \tilde{W}_{Y|Z}$ as
$\tilde{W}_{Y|Z}(y|z) W_Z(z|x) U_X(x)=
\tilde{W}_{YZ}(y,z|x)U_X(x)$.
Since the channels $W_Y,W_Z$ are conditional additive,
we have
$D_{h}^\varepsilon ( \tilde{W}_{YZ|X=x^n}^n  \| (\tilde{W}_{Y|Z}\times W_{Z|X=x^n})^n)
=D_{h}^\varepsilon ( \tilde{W}_{YZ|X=0}^n  \| (\tilde{W}_{Y|Z}\times W_{Z|X=0})^n)$
for $x^n \in {\cal X}^n$.
Hence, $
D_{h}^\varepsilon ( \tilde{W}_{YZ}^n\times P_{X^n} \| 
\tilde{W}_{Y|Z}^n\times W_{Z}^n\times P_{X^n})
=
D_{h}^\varepsilon ( \tilde{W}_{YZ|X=0}^n  \| (\tilde{W}_{Y|Z}\times W_{Z|X=0})^n)$.
We have
\begin{align}
D( \tilde{W}_{YZ|X=0}  \| \tilde{W}_{Y|Z}\times W_{Z|X=0})
=&
D (\tilde{W}_{YZ}\times P_{X} \| 
\tilde{W}_{Y|Z}\times W_{Z}\times P_{X})=I(X;Y|Z)_{\tilde{W}_{YZ}\times P_{X}} \nonumber \\
=& H(P_{X,\tilde{Z}}| P_{\tilde{Z}})
-H(P_{X,\tilde{Y}}| P_{\tilde{Y}}).
\Label{FOY}
\end{align}
Thus, we have the following theorem.

\begin{theorem}\Label{T13}
When a wire-tap channel $W_Y,W_Z$ is conditional additive
and degraded,
we have
\begin{align}
&\log N^{\varepsilon,\delta}(W_Y^n,W_Z^n)
\stackrel{(a)}{\le} 
D_{h}^{\varepsilon+\delta} ( \tilde{W}_{YZ|X=0}^n  \| (\tilde{W}_{Y|Z}\times W_{Z|X=0})^n) \nonumber \\
\stackrel{(b)}{=}&n (H(P_{X,\tilde{Z}}| P_{\tilde{Z}})
-H(P_{X,\tilde{Y}}| P_{\tilde{Y}})) +\sqrt{n}\sqrt{V(\tilde{W}_{YZ|X=0}\| \tilde{W}_{Y|Z}\times W_{Z|X=0})}\Phi^{-1}(\varepsilon+\delta) \nonumber \\
&+\frac{1}{2}\log n + F_{4}^{\varepsilon+\delta}(\tilde{W}_{YZ|X=0} \| \tilde{W}_{Y|Z}\times W_{Z|X=0})+
O(\frac{1}{\sqrt{n}}).
\Label{LGV}
\end{align}
\hfill $\square$\end{theorem}

\begin{IEEEproof}
The inequality $(a)$ of \eqref{LGV} follows from
Proposition \ref{P4-B}.
The second part $(b)$ of \eqref{LGV}
follows from 
\eqref{FOY} and Theorem \ref{T3}.
\end{IEEEproof}

The paper \cite[Theorem 13]{YSP} derived a similar evaluation as \eqref{LGV} up to the order of $\sqrt{n}$
in generic channels.
While the discussed more generic channels, they did not derived the constant term.

\subsection{Binary symmetric channels}
As an example of additive wire-tap channel, 
we consider the pair of binary symmetric channels,
in which, $W_Y$ and $W_Z$ are
the binary symmetric channels with crossover probability $p_Y$ and $p_Z$ with $\frac{1}{2}> p_Z>p_Y>0$
and ${\cal X}={\cal Y}={\cal Z}=\mathbb{F}_2$.
Hence, the sets $\tilde{\cal Y}$ and $\tilde{\cal Z}$ are trivial sets.
 Then, \eqref{LMY} is simplified to 
\begin{align}
&\log N^{\varepsilon,\delta}(W_Y^n,W_Z^n) \nonumber \\
\ge 
& n (h(p_Z)-h(p_Y))
 +\sqrt{n}
 (\sqrt{v(p_Y)}\Phi^{-1}(\varepsilon)+
 \sqrt{v(p_Z)}\Phi^{-1}(\delta)) 
 -\frac{1}{2}\log n
 + f_{5}^\varepsilon(p_Y)
+ f_{3}^\delta (p_Z)
  +O(\frac{1}{\sqrt{n}}),\Label{LLB1}
  \end{align}
where $h(p)$ is the binary entropy,
$v(p)$ is the varentropy,
and $f_{i}^\varepsilon (p)$ is $F_i^\varepsilon(P\|Q)$ when
$P$ is the binary distribution with flip probability $p$
and $Q$ is the identify function $1$.

Next, we proceed to the converse part.
The channel $\tilde{W}_{Z|Y}$ is the binary symmetric channel with the crossover probability
$\frac{p_Z-p_Y}{1-2p_Y} $, which is the solution of $p+p_Y- 2 p p_Y=p_Z$ with respect to $p$.
The channel $\tilde{W}_{Y|Z}$ is also the binary symmetric channel with the crossover probability $\frac{p_Z-p_Y}{1-2p_Y} $.
We define two distributions $P_{YZ}^1$ and $P_{YZ}^2$ as follows.
\begin{align}
P_{YZ}^1(0,0)&=(1-p_Y)(1-\frac{p_Z-p_Y}{1-2p_Y} ),\quad
P_{YZ}^1(0,1)=(1-p_Y) \frac{p_Z-p_Y}{1-2p_Y} , \\
P_{YZ}^1(1,0)&=p_Y(1-\frac{p_Z-p_Y}{1-2p_Y} ),\quad
P_{YZ}^1(1,1)=p_Y\frac{p_Z-p_Y}{1-2p_Y} .\\
P_{YZ}^2(0,0)&=(1-p_Z)(1-\frac{p_Z-p_Y}{1-2p_Y} ),\quad
P_{YZ}^2(0,1)=p_Z \frac{p_Z-p_Y}{1-2p_Y} , \\
P_{YZ}^2(1,0)&=(1-p_Z)(1-\frac{p_Z-p_Y}{1-2p_Y} ),\quad
P_{YZ}^2(1,1)=p_Z\frac{p_Z-p_Y}{1-2p_Y} .
\end{align}
Then, \eqref{LGV} is simplified to 
\begin{align}
&\log N^{\varepsilon,\delta}(W_Y^n,W_Z^n) \nonumber \\
\le 
&n (h(p_Z)-h(p_Y))
 +\sqrt{n}\sqrt{V(P_{YZ}^1\| P_{YZ}^2)}\Phi^{-1}(\varepsilon+\delta) \nonumber \\
&+\frac{1}{2}\log n + F_{4}^{\varepsilon+\delta}(P_{YZ}^1\| P_{YZ}^2)+
O(\frac{1}{\sqrt{n}}).
\Label{LGVX}
\end{align}

Fig. \ref{COM} numerically compares the upper and lower bounds in \eqref{LLB1} and \eqref{LGVX}.
Also, it compares with the second order approximations given in \cite{YSP}.
It shows that our higher correction is not so negligible.

\begin{figure}[h]
    \centering
    \includegraphics[width=0.98\hsize]{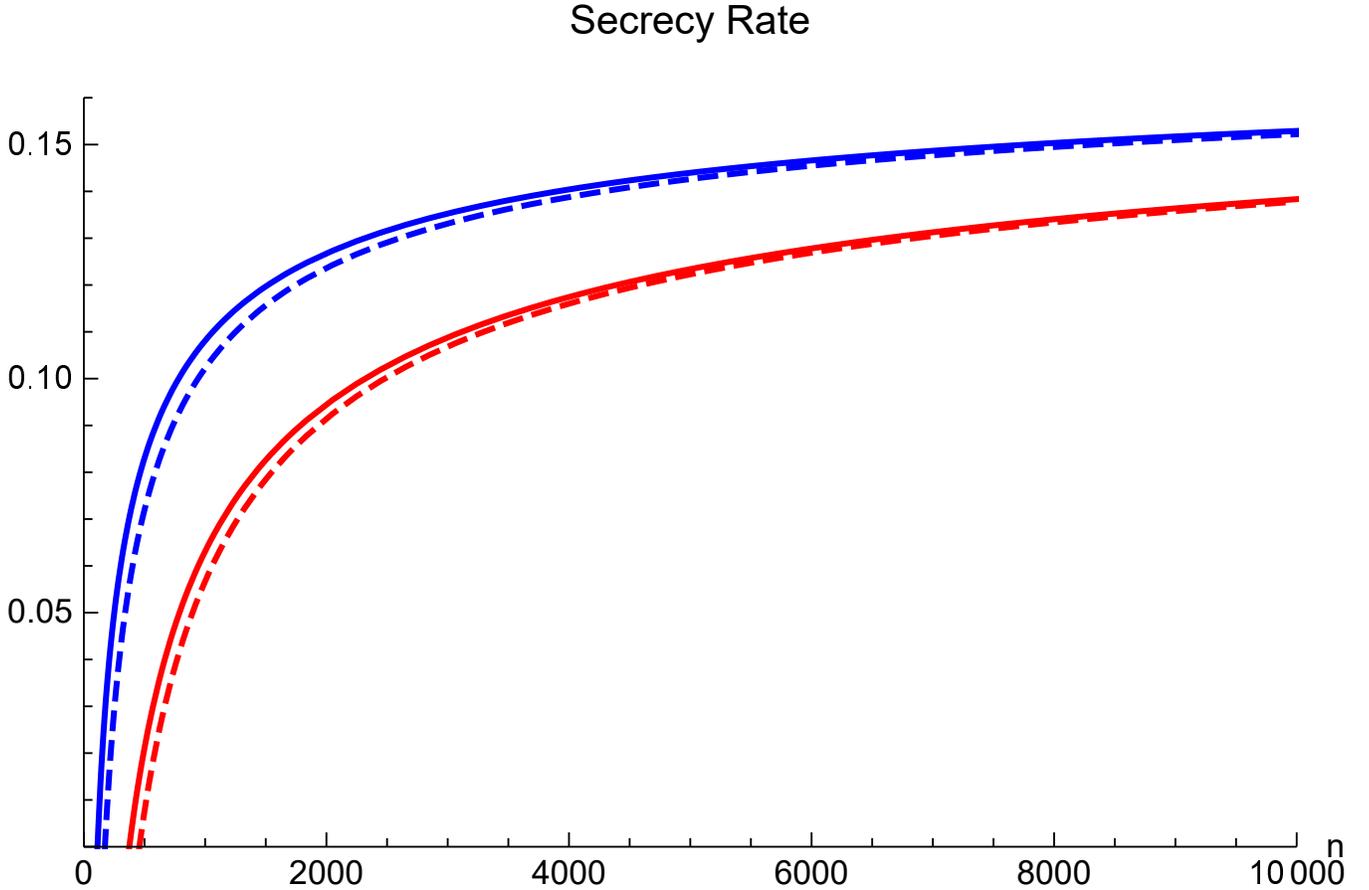}
    \caption{Upper and lower bounds of the transmission rate when $\epsilon=\delta=0.001$ and $P_Z=0.2$, $P_Y=0.1$. The horizontal axis expresses the block-length $n$, and the vertical axis expresses the transmission rate $\frac{1}{n}\log N^{\varepsilon,\delta}(W_Y^n,W_Z^n)$.
    Blue slid curve expresses the upper bound given in \eqref{LGVX}, and
    Red slid curve expresses the lower bound given in \eqref{LLB1}.
    Blue dashed curve expresses the second order approximation of the upper bound given in \eqref{LGVX}, and
    Red dashed curve expresses the second order approximation of the lower bound given in \eqref{LLB1}.
    The second order approximation means the value up to the order $\sqrt{n}$ in the respective terms.
    }
    \Label{COM}
\end{figure}

\subsection{BPSK scheme}
As an example of conditional additive wire-tap channel, 
we consider a pair of BPSK Gaussian channels.
As shown in \cite[Section IV-C]{Markov}, 
a BPSK Gaussian channel is conditional additive.
Our Gaussian wire-tap channel is given as
\begin{align}
Y=X+N_Y, Z=X+N_Z, 
\end{align}
where $N_Y$ and $N_Z$ are subject to the Gaussian distribution with average $0$
and variance $\sigma_Y^2$ and $\sigma_Z^2$, respectively.
We assume the relation $\sigma_Y^2<\sigma_Z^2$.
The input signal $X$ is limited to $1$ or $-1$.
Let 
$\varphi_{a,\sigma^2}$ be the probability density function of 
the Gaussian distribution with average $a$
and variance $\sigma^2$.
We define the distribution $\tilde{\varphi}_{\sigma} $
by $\tilde{\varphi}_{\sigma^2}(x):= 
\frac{1}{2}\varphi_{1,\sigma^2}(x)+ \frac{1}{2}\varphi_{-1,\sigma^2}(x)$.
The evaluation \eqref{LMY0} is rewritten as
\begin{align}
&\log N^{\varepsilon,\delta}(W_Y^n,W_Z^n) \nonumber \\
\ge 
& n (C(W_Y)-C(W_Z))
 +\sqrt{n}
 (\sqrt{V({\varphi}_{1,\sigma_Y^2} \| \tilde{\varphi}_{\sigma_Y^2})}
 \Phi^{-1}(\varepsilon)+
 \sqrt{V({\varphi}_{1,\sigma_Z^2} \| \tilde{\varphi}_{\sigma_Z^2})}\Phi^{-1}(\delta)) \nonumber \\
& -\frac{1}{2}\log n
 + F_{5}^\varepsilon ({\varphi}_{1,\sigma_Y^2} \| \tilde{\varphi}_{\sigma_Y^2})
+ F_{3}^\delta ({\varphi}_{1,\sigma_Z^2} \| \tilde{\varphi}_{\sigma_Z^2})
  +O(\frac{1}{\sqrt{n}}) .\Label{LMY3}
\end{align}

Next, we consider the converse part.
The channel $\tilde{W}_{Z|Y}$ is given as
\begin{align}
Z= Y+ N_{Z|Y},
\end{align}
where $N_{Z|Y}$ is subject to the Gaussian distribution with average $0$
and variance $\sigma_Z^2-\sigma_Y^2$.
%However, The channel $\tilde{W}_{Y|Z}$ is not so simple.
Hence, 
the probability density function of the joint conditional distribution 
$p_{YZ|X}(y,z|x)$ is given as $p_{YZ}(y,z)= 
{\varphi}_{x,\sigma_Y^2}(y) \varphi_{y,\sigma_Z^2}(z) $.
The probability density function of the joint distribution $p_{YZ}(y,z)$
is given as $p_{YZ}(y,z)= 
\tilde{\varphi}_{\sigma_Y^2}(y) \varphi_{y,\sigma_Z^2}(z) $.
The probability density function of the conditional distribution $p_{Y|Z}(y|z)$
is given as $p_{Y|Z}(y|z)= 
\tilde{\varphi}_{\sigma_Y^2}(y) \varphi_{y,\sigma_Z^2}(z) 
\tilde{\varphi}_{\sigma_Z^2}(z)^{-1} $.
Therefore, \eqref{LGV} is simplified to
\begin{align}
&\log N^{\varepsilon,\delta}(W_Y^n,W_Z^n) \nonumber \\
\le 
&n (C(W_Y)-C(W_Z))
  +\sqrt{n}
  \sqrt{V ( p_{YZ|X=0} \| p_{Y|Z}\times \varphi_{0,\sigma_Z^2} )}
\Phi^{-1}(\varepsilon+\delta) \nonumber \\
&+\frac{1}{2}\log n + F_{4}^{\varepsilon+\delta}
( p_{YZ|X=0} \| p_{Y|Z}\times \varphi_{0,\sigma_Z^2} )
+
O(\frac{1}{\sqrt{n}}).
\Label{LGV9}
\end{align}

The paper \cite[Theorem 14]{YSP} discussed the Gaussian channel with energy constraint.
This paper focuses on BPSK scheme, which is different from the result of \cite[Theorem 14]{YSP}.

\subsection{Secure communication based on correlated random variable}
As an example of conditional additive wire-tap channel, 
we consider secure communication by the correlated variables 
$\tilde{X},\tilde{Y}$, and $\tilde{Z}$, which 
are subject to the joint distribution $P_{\tilde{X},\tilde{Y},\tilde{Z}}$,
and take values in 
$\tilde{{\cal X}},\tilde{\cal Y}$, $\tilde{\cal Z}$, respectively \cite[Section VI]{hayashi:10}.
Assume that the sender,
the legitimate receiver, and the eavesdropper have
the variables $\tilde{X}$,
$\tilde{Y}$, and $\tilde{Z}$, respectively.
We assume that $\tilde{\cal X}$ equals an additive group ${\cal X}$.

Then, we consider the wire-tap channel with the input system ${\cal X}$ in the following way.
When the input is $X\in {\cal X}$, the sender sends $X+\tilde{X}$
to the legitimate receiver via a public channel.
Hence, due to the property of the public channel,
the information $X+\tilde{X}$ is also leaked to the eavesdropper.
That is, the legitimate receiver receives the variable $Y=(\tilde{Y},X+\tilde{X})
\in {\cal Y}:= {\cal X} \times \tilde{\cal Y}$,
and 
the eavesdropper receives the variable $Z=(\tilde{Z},X+\tilde{X})
\in {\cal Z}:= {\cal X} \times \tilde{\cal Z}$.
The conditional distribution to characterize our wire-tap channel is 
given as 
$W_Y(\tilde{x},\tilde{y}|x)=
P_{\tilde{X},\tilde{Y}}(\tilde{x}-x,\tilde{y})$
and
$W_Z(\tilde{x},\tilde{z}|x)=
P_{\tilde{X},\tilde{Z}}(\tilde{x}-x,\tilde{z})$.
Hence, the wire-tap channel $W_Y,W_Z$ is conditional additive.
In this model, 
we have $P_{X,\tilde{Y}}=P_{\tilde{X},\tilde{Y}} $ and $P_{X,\tilde{Z}}=P_{\tilde{X},\tilde{Z}} $.
Hence, we obtain a lower bound of 
$\log N_{\varepsilon,\delta,n,a}(W_Y,W_Z)$ of this model
by replacing $P_{X,\tilde{Y}}$ and $P_{X,\tilde{Z}}$
by $P_{\tilde{X},\tilde{Y}} $ and $P_{\tilde{X},\tilde{Z}} $
in the RHS of \eqref{LMY}.

Next, we consider the converse part.
For this aim, we assume the Markovian chain $\tilde{X}-\tilde{Y}-\tilde{Z}$. 
Using
$W_{Z|Y}( \tilde{x},\tilde{z}|\tilde{x},\tilde{y})
:= \delta_{\tilde{x},\tilde{x}'} P_{\tilde{Z}|\tilde{Y}}(\tilde{z}|\tilde{y})$,
we have $W_{Z|Y}\cdot W_{Y|X}=W_{Z|X}$.
Hence, this wire-tap channel is degraded.
Also, $\tilde{W}_{Y|Z}$ is calculated as
$\tilde{W}_{Y|Z}( \tilde{x},\tilde{y}|\tilde{x},\tilde{z})
= \delta_{\tilde{x},\tilde{x}'} P_{\tilde{Y}|\tilde{Z}}(\tilde{y}|\tilde{z})$.
Since $W_{YZ}( \tilde{x},\tilde{y},\tilde{x}',\tilde{z}|x  )
=\delta_{\tilde{x},\tilde{x}'} P_{\tilde{X},\tilde{Y},\tilde{Z}}(\tilde{x}-x, \tilde{y},\tilde{z})$, we have 
\begin{align}
W_{YZ|X=0}( \tilde{x},\tilde{y},\tilde{x}',\tilde{z})
&=\delta_{\tilde{x},\tilde{x}'} P_{\tilde{X},\tilde{Y},\tilde{Z}}(\tilde{x}, \tilde{y},\tilde{z}) \\
\tilde{W}_{Y|Z}\times W_{Z|X=0}
( \tilde{x},\tilde{y},\tilde{x}',\tilde{z})
&= \delta_{\tilde{x},\tilde{x}'} P_{\tilde{Y}|\tilde{Z}}(\tilde{y}|\tilde{z})
P_{\tilde{X},\tilde{Z}}(\tilde{x}',\tilde{z}).
\end{align}
Hence,
the upper bound of 
$\log N^{\varepsilon,\delta}(W_Y^n,W_Z^n) $ 
given in the RHS of \eqref{LGV} can be calculated to 
\begin{align}
&n (H(P_{\tilde{X},\tilde{Z}}| P_{\tilde{Z}})
-H(P_{\tilde{X},\tilde{Y}}| P_{\tilde{Y}})) 
+\sqrt{n}\sqrt{V(P_{\tilde{X},\tilde{Y},\tilde{Z}} \| 
P_{\tilde{Y}|\tilde{Z}} \times P_{\tilde{X},\tilde{Z}}
)}\Phi^{-1}(\varepsilon+\delta) \nonumber \\
&+\frac{1}{2}\log n + F_{4}^\varepsilon(P_{\tilde{X},\tilde{Y},\tilde{Z}} \| 
P_{\tilde{Y}|\tilde{Z}} \times P_{\tilde{X},\tilde{Z}}
)+
O(\frac{1}{\sqrt{n}}).
\Label{LGV2}
\end{align}

\section{Strong large deviation and Edgeworth expansion}\Label{S7}
Let $p$ be a non-negative measure and
$d_S$ be the lattice span of the real valued function $X$, 
which is defined as follows.
Let $S$ be the set of the support of the measure $p \circ X^{-1}$.
When there exists a non-negative value $x$ satisfying 
$ \{ a-b\}_{a,b\in S}  \subset x \bZ$,
the real valued function $X$ is called a lattice function or a lattice variable.
Then, 
the lattice span $d_S$ is defined as the maximum value of the above non-negative value $x$.
Denoting all of elements of $S$ as $a_1<a_2< \ldots < a_l$,
we have 
\begin{align}
d_S= \min_{n_i \in \bZ} \bigg\{  \sum_{i=1}^l n_i a_i \Bigg| \sum_{i=1}^l n_i =0 ,~ 
\sum_{i=1}^l n_i a_i>0\bigg\}
\end{align}
due to the following reason;
When integers $y_1, \ldots, y_l$ have the greatest common divisor $1$,
there exist integers $n_1, \ldots, n_l$ such that $\sum_{i=1}^l n_i y_i=1$.
When there does not exist such a non-negative value $x$,
the real valued function $X$ is called 
a non-lattice function or a non-lattice variable.
Then, the lattice span $d_S$ is regarded as zero.
Now, we summarize the fundamental properties for the lattice and non-lattice cases.
For this purpose,
we denote the set $ \{\sum_{i=1}^n a_i \}_{a_i \in S}$ by $S_n$.

\begin{lemma}\Label{L9-20}
%We denote all of elements of $S$ as $a_1<a_2< \ldots < a_l$.
%Let $a_{}$ and $a_{\max}$ be the minimum and maximum of $S$.
We fix a small real number $\delta>0$.
In the lattice case, 
there exists a sufficiently large integer $N$ 
such that $S_n$ satisfies the following condition for any $n \ge N$. 
Denote all of elements of $S_n\cap [n(a_1+\delta),n(a_l-\delta)]$
as $b_1<b_2< \ldots < b_k$.
We have $b_{i+1}-b_i= d_S$.

In the non-lattice case, 
for an arbitrary  small real number $\varepsilon$,
there exists a sufficiently large integer $N$ 
such that $S_n$ satisfies the following condition for any $n \ge N$. 
Denote all of elements of $S_n\cap [n(a_1+\delta),n(a_l-\delta)]$
as $b_1<b_2< \ldots < b_k$.
We have $b_{i+1}-b_i< \varepsilon$.
\hfill $\square$\end{lemma}

\begin{IEEEproof}
\PF{Lattice case}
Since the definition of $d_S$ guarantees that $b_{i+1}-b_i\ge d_S$,
it is enough to show that $b_{i+1}-b_i\le d_S$.
Assume that integers $n_i$ satisfies the equations 
\begin{align}
\sum_{i=1}^l n_i a_i &=d_S \Label{9-20-1} \\
\sum_{i=1}^l n_i &=0.\Label{9-20-2}
\end{align}
We define the subsets $S_+:= \{a_i \in S| n_i \ge 0 \}$ and
$S_-:= \{a_i \in S| n_i < 0 \}$,
the positive integers
$m_2:= \sum_{i:a_i\in S_+}n_i$ and $m_1:= (a_l-a_1)/d_S$,
and the positive real numbers
$A:= -m_1 \sum_{i :a_i\in S_-}n_i a_i$,
$B:= m_1 \sum_{i :a_i\in S_+}n_i a_i$,
$\delta_-:= (A-a_1 m_1 m_2)/n$, and
$\delta_+:= (a_l m_1 m_2-B+ m_1d_s)/n$.

So, we have $n (a_1+\delta_-)= a_1 (n-m_1 m_2)+A= n a_1+ (A-a_1 m_1 m_2)$
and $n (a_l-\delta_+)= a_l (n-m_1 m_2)+B= n a_l- (a_l m_1 m_2-B)$.
We choose an element $x:=n (a_1+\delta_-) + (c_1 m_1+c_2)d_S
\in [n (a_1+\delta_-),n (a_l-\delta_+)]$ with 
integers $c_1$ and $c_2 \le m_1$.
When $(c_1 m_1+c_2)$ takes the maximum, $x$ is $n (a_l-\delta_+)$, i.e.,
$c_1 m_1+c_2= (n-m_1 m_2) m_1$.
So, the maximum of $c_1$ is $n-m_1 m_2$.

Using \eqref{9-20-1} and the definitions of $\delta_-$ an $A$, 
we have 
\begin{align}
x=& c_1 a_l+(n-c_1-m_1 m_2) a_1
+      c_2\Big(\sum_{i:a_i\in S_+}n_i a_i\Big) 
\nonumber \\
&- (m_1-c_2)\sum_{i:a_i\in S_-}n_i a_i \stackrel{(a)}{\in} S_n.
\end{align}
Here, the relation $(a)$ follows from the following facts;
$c_1$ and $ (n-c_1-m_1 m_2)$ are non-negative integers, 
$c_2 n_i$ is a non-negative integer for $i \in S_+$,
and $ - (m_1-c_2)n_i$ is a non-negative integer for $i \in S_-$. 
Thus, when we denote all of elements of $S_n\cap [n (a_1+\delta_-),n (a_l-\delta_+)]$ as $b_1<b_2< \ldots < b_k$.
We have $b_{i+1}-b_i \le d_S$.
When $n$ is sufficiently large, we have $\delta_-, \delta_+ \le \delta$.
So, we obtain the desired statement.

\PF{Non-lattice case}
For an arbitrary $\varepsilon>0$, we can take integers $n_i$ such that
$0<\tilde{d}:=\sum_{i=1}^l n_i a_i< \varepsilon$ and $\sum_{i=1}^l n_i=0$.
(If impossible, we have the minimum of $\sum_{i=1}^l n_i a_i$ with $\sum_{i=1}^l n_i=0$
is strictly larger than $0$, which contradicts $d_S=0$.)
We redefine $m_1:= \lceil(a_l-a_1)/\varepsilon\rceil$, 
and define other terms in the same way by replacing $d_S$ by $\tilde{d}$.
Using the same discussion, we find that 
the element $x:=n (a_1+\delta_-) + c_1 (a_l-a_1)+ c_2 \tilde{d}
\in [n (a_1+\delta_-),n (a_l-\delta_+)]$ with $c_2 \le m_1$
belongs to $S_n$.
When $n$ is sufficiently large, we have $\delta_-, \delta_+ \le \delta$.
So, we have $b_{i+1}-b_i< \varepsilon$.
\end{IEEEproof}

Here $p$ is not necessarily normalized.
Define the notation 
$\mathbb{E}_p [X] \stackrel{\rm def}{=} \int X(\omega) p(d \omega)$.
Define the cumulant generating function 
$\tau (s)\stackrel{\rm def}{=} \log \mathbb{E}_p [e^{sX}]$.
Denote the inverse function of the derivative $\tau' (s) $ by $\eta$.

\begin{proposition}[\protect{Bahadur and Rao \cite{BR}, \cite[Theorem 3.7.4]{Dembo98}}]\Label{11-4-4}
Assume that $\tau (0)<\infty $.
When $R > \frac{\mathbb{E}_p[X]}{\mathbb{E}_p[1]}$, 
we have
\begin{align}
&\log p^n\{ X_n \ge n R\} \nonumber \\
=& \chi_0(R) n - \frac{1}{2}\log n+\chi_1(R)+\chi_2(R)\frac{1}{n}+o(\frac{1}{n}), \\
&\log p^n\{ X_n \le n R\}
= n \tau (0) +o(1) ,
\end{align}
where
\begin{align}
\chi_0(R)&\stackrel{\rm def}{=}-R \eta(R) + \tau (\eta (R)) 
\end{align}
and
\begin{align}
& \chi_1(R) \nonumber \\
\stackrel{\rm def}{=} &
\left\{
\begin{array}{ll}
-\frac{1}{2}\log 2 \pi -\log \eta(R) + \frac{1}{2}\log \eta'(R)
& \hbox{ if } d_S=0 \\
-\frac{1}{2}\log 2 \pi + \frac{1}{2}\log \eta'(R)
+ \log \frac{d_S}{1-e^{-d_S \eta(R)}} 
& \hbox{ if } d_S>0,
\end{array}
\right.
\end{align}
and $\chi_2(R)$ is a continuous function.
The convergences of the differences between the LHSs and RHSs 
are compact uniform.
\hfill $\square$\end{proposition}

As a generalization of the function $v(d)$, we define
the function
$v(d,s)$ as
\begin{align}
v(d,s):= 
\left\{
\begin{array}{ll}
\log \frac{d}{1-e^{-ds}} & \hbox{ when } d > 0 \\
-\log s & \hbox{ when } d = 0 .
\end{array}
\right.
\end{align}
Using this function, we have the following lemma.

\begin{lemma}\Label{L2}
Assume that $\tau'(s_0)=R_0 $ and $s_0 >0$.
When $R= R_0+\frac{R_1}{\sqrt{n}}+\frac{R_2}{n}$, we have
\begin{align}
&\log p^n\{ X_n \ge n R\} \nonumber \\
=& n (- R s_0+\tau(s_0))-\frac{1}{2}\log (2\pi \tau''(s_0) n )
+ v(d_S,s_0)
- \frac{R_1^2}{2\tau''(s_0)} 
+O( \frac{1}{\sqrt{n}}).\Label{KKY3}
\end{align}
\hfill $\square$\end{lemma}

\begin{IEEEproof}
Let $s$ be the real number to satisfy $\tau'(s)=R$.
Since $\tau'(s)=\tau'(s_0)+ \tau''(s_0)(s-s_0) +O((s-s_0)^2) $,
we have $s-s_0= \frac{R-R_0}{\tau''(s_0)}+O((s-s_0)^2)$.
Thus,
\begin{align}
\chi_0(R)=&-R s + \tau (s)
 =-R s + \tau (s_0)+\tau' (s_0)(s-s_0)
+\frac{1}{2}\tau'' (s_0)(s-s_0)^2 +O((s-s_0)^3)\nonumber \\
 =&-R s_0+ \tau (s_0)-R( s-s_0) 
 +\tau' (s_0)(s-s_0)
+\frac{1}{2}\tau'' (s_0)(s-s_0)^2 +O((s-s_0)^3)\nonumber \\
 =&-R s_0+ \tau (s_0)-(R-R_0)( s-s_0) 
+\frac{1}{2}\tau'' (s_0)(s-s_0)^2 +O((s-s_0)^3)\nonumber \\
 =&-R s_0+ \tau (s_0)
 -\frac{(R-R_0)^2}{2\tau'' (s_0)} +O((s-s_0)^3) ,\Label{KKY}
\end{align}
and
 \begin{align}
\eta'(R)= \frac{1}{\tau''(s)}
= \frac{1}{\tau''(s_0)}+ O(s-s_0).
\end{align}

In the non-lattice case, we have
\begin{align}
& \chi_1(R) \nonumber \\
= &
-\frac{1}{2}\log 2 \pi -\log \eta(R) + \frac{1}{2}\log \eta'(R)
= 
-\frac{1}{2}\log 2 \pi -\log s + \frac{1}{2}\log \eta'(R) \nonumber\\
= &
-\frac{1}{2}\log 2 \pi -\log s_0 - \frac{1}{2}\log  \tau''(s_0)+ O(s-s_0).\Label{KKY2}
\end{align}
Since $s-s_0=O(\frac{1}{\sqrt{n}}) $ and $R-R_0= \frac{R_1}{\sqrt{n}}+O(\frac{1}{n})$, 
the combination of Proposition \ref{11-4-4}, \eqref{KKY}, and \eqref{KKY2} implies \eqref{KKY3}. 

In the lattice case, we replace $-\log \eta(R)$ by
$\log \frac{d_S}{1-e^{-d_S \eta(R)}}$.
This value is calculated to be $ v(d_S,s_0)+O(\frac{1}{\sqrt{n}})$. 
Hence, we obtain \eqref{KKY3}. 
\end{IEEEproof}

When $R $ is close to $\tau'(0) $, we have the Gaussian approximation.
In this case, we have Edgeworth expansion instead of the strong large deviation.

\begin{proposition}[\protect{\cite{BG}}]\Label{L1}
Assume that $p$ is a probability distribution. % and $\tau$ is a smooth function.
Define the skewness, i.e., the normalized version of the third cumlant  
\begin{align}
\kappa:= \mathbb{E}_{p} \Big[ \Big(\frac{X-\mathbb{E}_{p}[X]}
{\sqrt{\mathbb{V}_{p}[X]}}\Big)^3\Big]= \frac{\tau'''(0)}{\tau''(0)^{3/2}}.
\end{align}
Then, we have
\begin{align}
p^n\{ X_n \le n E[X] + \sqrt{n}\sqrt{V[X]}x \}
=\Phi(x)-\varphi(x)\frac{\kappa (x^2-1)}{6}\frac{1}{\sqrt{n}}+O(\frac{1}{n}),
\end{align}
where $\varphi $ is the probability density function of the standard Gaussian distribution.
\hfill $\square$\end{proposition}

\section{Proof of Theorem \ref{T2}}\Label{S8}
In this section, we abbreviate $d_{(P_{AE}\|P_E)} $ to $d$.
To prove the relation \eqref{GS1}, we prepare the following Lemma.
\begin{lemma}\Label{L01}
For $P_{AE}\in {\cal P}({\cal A}\times {\cal E})$, we have
\begin{align}
&{\delta}_{\min}(m|P_{AE}|P_E)=
P_{AE}
\Big\{ -\log \frac{P_{AE}(a,e)}{P_E(e)} \le  m \Big \}
-e^{-m}P_E \Big\{ -\log \frac{P_{AE}(a,e)}{P_E(e)} \le  m \Big \}.
\Label{E14Y}
\end{align}
\hfill $\square$\end{lemma}
\begin{IEEEproof}
First, we use another expression of $\delta_{\min}(m | P_{AE}|P_E)$ as
\begin{align*}
\delta_{\min}(m | P_{AE}|P_E)= 
\min_{Q_{AE} \in {\overline{\cal P}}({\cal A} \times {\cal E})}
\{ 2d(Q_{AE}, P_{AE})|  H_{\min}(Q_{AE}|P_E) \le m \}.
\end{align*}
Th optimal $Q_{AE,opt} \in {\overline{\cal P}}({\cal A} \times {\cal E})$
is given as follows.
\begin{align}
Q_{AE,opt}(a,e):=
\left\{
\begin{array}{ll}
e^{-m}P_E(e) & \hbox{ when }  -\log \frac{P_{AE}(a,e)}{P_E(e)} \le  m \\
P_{AE}(a,e)& \hbox{ when }  -\log \frac{P_{AE}(a,e)}{P_E(e)} >  m .
\end{array}
\right.
\end{align}
Since $2d(Q_{AE,opt}, P_{AE})$ equals to the RHS of \eqref{E14Y},
we obtain Eq. \eqref{E14Y}.
\end{IEEEproof}

The relation \eqref{GS1} follows from the following Lemma \ref{L6X}.
\begin{lemma}\Label{L6X}
For $P_{AE}\in {\cal P}({\cal A}\times {\cal E})$, we have
\begin{align}
{H}_{\min}^\varepsilon(P_{AE}^n|P_E^n)
=nH(P_{AE}|P_E) +\sqrt{n}\sqrt{V(P_{AE}\|P_E)}\Phi^{-1}(\varepsilon)
+ F_{1}^\varepsilon(P_{AE}\|P_E)+O(\frac{1}{\sqrt{n}}).
\end{align}
%where
%\begin{align}
%F_{1}^\varepsilon(P_{AE}\|P_E):=
%\sqrt{V(P_{AE}\|P_E)} \Big(\frac{\kappa(P_{AE}\|P_E)(\Phi^{-1}(\varepsilon)^2-1)}{6 }-
%e^{ \frac{1}{2V(P_{AE}\|P_E)}+ \Phi^{-1}(\varepsilon)^2} \Big).
%\end{align}
\hfill $\square$\end{lemma}

%$:= E \Big[ \Big(\frac{X-E[X]}{\sqrt{V[X]}}\Big)^3\Big]$,

\begin{IEEEproof}
We choose $m:=nH(P_{AE}|P_E) +\sqrt{n}\sqrt{V(P_{AE}\|P_E)}B_1+ \sqrt{V(P_{AE}\|P_E)} B_2$ with $B_1:= \Phi^{-1}(\varepsilon)$.
We apply Proposition \ref{L1} to the case with $X=-\log \frac{P_{AE}}{P_E} $ and the distribution $P_{AE} $.
Then, we obtain
\begin{align}
&P_{AE}^n
\Big\{ -\log \frac{P_{AE}^n(a,e)}{P_E^n(e)} \le  m \Big \}
\nonumber \\
=& \Phi(B_1) -\varphi(B_1)\frac{\kappa(P_{AE}\|P_E)(B_1^2-1)}{6 \sqrt{n}}
+\varphi(B_1) \frac{B_2}{\sqrt{n}}
+O(\frac{1}{n}) \nonumber \\
=& \varepsilon
+\frac{1}{\sqrt{2\pi n}}e^{- B_1^2/2}
(B_2-\frac{\kappa(P_{AE}\|P_E)(B_1^2-1)}{6 }
)
+O(\frac{1}{n}) \Label{E12}.
\end{align}
Next, we apply Lemma \ref{L2} to the case with 
the measure $P_{E} $ on ${\cal A} \times {\cal E}$, 
$X=\log \frac{P_{AE}}{P_E} $, 
$R=- \frac{m}{n}$,
$s_0=1$, and $R_0=-H(P_{AE}\|P_E)$.
Then, we obtain
\begin{align}
&
\log \Big[e^{-m}P_E \Big\{ -\log \frac{P_{AE}^n(a,e)}{P_E^n(e)} \le  m \Big \} \Big]
\nonumber \\
=&
-\frac{1}{2}\log (2 \pi V(P_{AE}\|P_E) n) 
-\frac{B_1^2}{2} +v(d) 
+O(\frac{1}{\sqrt{n}}).
\Label{E13}
\end{align}
Hence, we have
\begin{align}
&{\delta}_{\min}(m|P_{AE}^n|P_E^n)\stackrel{(a)}{=}
P_{AE}^n
\Big\{ -\log \frac{P_{AE}^n(a,e)}{P_E^n(e)} \le  m \Big \}
-e^{-m}P_E \Big\{ -\log \frac{P_{AE}^n(a,e)}{P_E^n(e)} \le  m \Big \} \nonumber \\
\stackrel{(b)}{=}& \varepsilon +\frac{1}{\sqrt{2\pi n}}e^{- B_1^2/2}
(B_2-\frac{\kappa(P_{AE}\|P_E)(B_1^2-1)}{6 }
)
- \frac{1}{\sqrt{2\pi V(P_{AE}\|P_E) n}}
e^{- \frac{1}{2} B_1^2+v(d)} +O(\frac{1}{{n}}),\Label{E14}
\end{align}
where
$(a)$ follows from Lemma \ref{L01}, and $(b)$ follows from
the combination \eqref{E12} and \eqref{E13}.
The equation ${\delta}_{\min}(m|P_{AE}^n|P_E^n)=\varepsilon +O(\frac{1}{n})$ holds
if and only if the following relation holds
%$ B_1= \Phi^{-1}(\varepsilon)$
\begin{align}
\log  (B_2-\frac{\kappa(P_{AE}\|P_E)(B_1^2-1)}{6 })
=-\frac{1}{2}\log V(P_{AE}\|P_E)+v(d),
\end{align}
which implies
\begin{align}
B_2=\frac{\kappa(P_{AE}\|P_E)(B_1^2-1)}{6 }+
\frac{1}{\sqrt{V(P_{AE}\|P_E)}}e^{v(d)}.
\end{align}
Therefore, the equation ${\delta}_{\min}(m|P_{AE}^n|P_E^n)=\varepsilon +O(\frac{1}{n})$ holds
if and only if 
$\sqrt{V(P_{AE}\|P_E)} B_2=
 F_{1}^\varepsilon(P_{AE}\|P_E)$.
This statement is equivalent to the desired statement.
\end{IEEEproof}

For the proofs of \eqref{GS2} and \eqref{GS3},
we prepare the following two lemmas.
\begin{lemma}\Label{L3}
The maximum $\max_{x}x-ae^{x/2}$ equals $2(\log 2 - \log a-1)$
and it is achieved by $x=2\log \frac{2}{a}$. 
The maximum $\max_{x}x-ae^{x}$ equals $-\log a-1$
and it is achieved by $x=-\log a$. 
\hfill $\square$\end{lemma}

\begin{lemma}\Label{L3B}
For $P_{AE}\in {\cal P}({\cal A}\times {\cal E})$, we have
\begin{align}
\ell_{\min}^{\varepsilon}(P_{AE}|P_E) = 
\max\{m'|
\exists m \hbox{ such that }
{\delta}_{\min}(m| P_{AE}|P_E)+ \frac{1}{2}e^{\frac{m'-m}{2}}
 \le \varepsilon\} 
\Label{GFT4}.
\end{align}
\hfill $\square$\end{lemma}
\begin{IEEEproof}
We have
\begin{align}
\delta_{\min}(m'|P_{AE}|P_E)=2\bar{\delta}_{\min}(m'|P_{AE}|P_E)
\Label{GFT}.
\end{align}
Also, we have
\begin{align}
\Delta_{\min}(m'|P_{AE}|P_E)
&= 
\min_{Q_{AE}\in \bar{{\cal P}}({\cal A} \times {\cal E})}
2 d(Q_{AE},P_{AE})
 + \frac{1}{2} \sqrt{e^{m'- {H}_{\min}(Q_{AE}|P_E)}} \nonumber \\
&=\min_{m}
2\bar{\delta}_{\min}(m | P_{AE}|P_E)+ \frac{1}{2}e^{\frac{m'-m}{2}}
\Label{GFT2}.
\end{align}
Combining \eqref{GFT} and \eqref{GFT2},
we have 
\begin{align}
\Delta_{\min}(m'|P_{AE}|P_E)
=\min_{m}
{\delta}_{\min}(m| P_{AE}|P_E)+ \frac{1}{2}e^{\frac{m'-m}{2}}
\Label{GFT3}.
\end{align}
Hence, we obtain \eqref{GFT4}.
\end{IEEEproof}

The relation \eqref{GS2} follows from the following Lemma \ref{L7}.

\begin{lemma}\Label{L7}
For $P_{AE}\in {\cal P}({\cal A}\times {\cal E})$, we have
\begin{align}
{\ell}_{\min}^\varepsilon(P_{AE}^n|P_E^n)
=nH(P_{AE}|P_E) +\sqrt{n}\sqrt{V(P_{AE}\|P_E)}\Phi^{-1}(\varepsilon)-\log n+ F_{2}^\varepsilon(P_{AE}\|P_E)+O(\frac{1}{\sqrt{n}}).
\Label{E17}
\end{align}
%where
%\begin{align}
%F_{2}^\varepsilon(P_{AE}\|P_E):=
%\sqrt{V(P_{AE}\|P_E)} \Big(\frac{\kappa(P_{AE}\|P_E)(\Phi^{-1}(\varepsilon)^2-1)}{6 }-
%e^{ \frac{1}{2V(P_{AE}\|P_E)}+ \Phi^{-1}(\varepsilon)^2} \Big)+2 (\log 2-2) - \log V(P_{AE}\|P_E) + \Phi^{-1}(\varepsilon)^2
%\end{align}
\hfill $\square$\end{lemma}

%$:= E \Big[ \Big(\frac{X-E[X]}{\sqrt{V[X]}}\Big)^3\Big]$,

\begin{IEEEproof}
In this proof, we employ the expression of $\ell_{\min}^{\varepsilon}(P_{AE}|P_E)$ given in Lemma \ref{L3B}.
We choose
$m:=nH(P_{AE}|P_E) +\sqrt{n}\sqrt{V(P_{AE}\|P_E)}B_1+ \sqrt{V(P_{AE}\|P_E)} B_2$
and
$m':=m+\log 2- \log (\pi n) + B_3$ with $B_1=\Phi^{-1}(\varepsilon)$.
Using \eqref{E14}, we have
\begin{align}
&({\delta}_{\min}(m|P_{AE}^n|P_E^n)+\frac{1}{2}e^{\frac{m'-m}{2}} )-\varepsilon\nonumber\\
=&
\frac{1}{\sqrt{2\pi n}}e^{- B_1^2/2}
(B_2-\frac{\kappa(P_{AE}\|P_E)(B_1^2-1)}{6 })
+O(\frac{1}{n}) 
-\frac{1}{\sqrt{2\pi V(P_{AE}\|P_E) n}}
e^{-\frac{1}{2} B_1^2+v(d)}
+O(\frac{1}{n}) 
+\frac{1}{\sqrt{2\pi n}}e^{\frac{B_3}{2}} \nonumber \\
=&
\frac{1}{\sqrt{2\pi n}}
(e^{- B_1^2/2}
(B_2-\frac{\kappa(P_{AE}\|P_E)(B_1^2-1)}{6 })
-\frac{1}{\sqrt{V(P_{AE}\|P_E)}} e^{-\frac{1}{2} B_1^2+v(d)}
+e^{\frac{1}{2}B_3} )
+O(\frac{1}{n}) .\Label{E14B}
\end{align}
The relation
\begin{align}
e^{- B_1^2/2}
(B_2-\frac{\kappa(P_{AE}\|P_E)(B_1^2-1)}{6 })
-\frac{1}{\sqrt{V(P_{AE}\|P_E)}}e^{-\frac{1}{2}B_1^2+v(d)}
+e^{\frac{1}{2}B_3} =0
\end{align}
holds if and only if
\begin{align}
B_2=B_{2,\min}(B_3):= \frac{\kappa(P_{AE}\|P_E)(B_1^2-1)}{6 }+
\frac{1}{\sqrt{V(P_{AE}\|P_E)}}e^{v(d)}
-e^{\frac{1}{2}(B_3+B_1^2)} .\Label{E15}
\end{align}
Applying Lemma \ref{L3}, we have
\begin{align}
&\sqrt{V(P_{AE}\|P_E)} \max_{B_3}B_{2,\min}(B_3)+B_3\nonumber \\
=&\max_{B_3}\sqrt{V(P_{AE}\|P_E)} (
\frac{\kappa(P_{AE}\|P_E)(B_1^2-1)}{6 }+
\frac{1}{\sqrt{V(P_{AE}\|P_E)}}e^{v(d)}
-e^{\frac{1}{2}(B_3+B_1^2)})+B_3 \nonumber \\
=&\sqrt{V(P_{AE}\|P_E)} (
\frac{\kappa(P_{AE}\|P_E)(B_1^2-1)}{6 }+
\frac{1}{\sqrt{V(P_{AE}\|P_E)}}e^{v(d)})
+2 \log 2 - \log V(P_{AE}\|P_E) - B_1^2-2.\Label{E16}
\end{align}
Due to the combination of \eqref{E14B}, \eqref{E15}, \eqref{E16}, and Lemma \ref{L3B},
when $B_2$ is chosen in \eqref{E15} and $B_3$ is chosen to achieve the maximum in \eqref{E16}, 
the value $m'$ equals to the RHS of \eqref{E17}
because the sum of the RHS of \eqref{E16} and $\log 2- \log \pi $
equals $F_{2}^\varepsilon(P_{AE}\|P_E)$.
Hence, we obtain \eqref{E17}.
\end{IEEEproof}

To prove the relation \eqref{GS3}, we prepare the following lemma.
\begin{lemma}\Label{L5X}
For $P_{AE}\in {\cal P}({\cal A}\times {\cal E})$, we have
\begin{align}
\Delta_{2}(m'|P_{AE}|P_E)
=& \min_{f}
\mathbb{E}_{P_{AE}}[1-f]
+\frac{1}{2}\sqrt{2^{m'}
\mathbb{E}_{\frac{P_{AE}^2}{P_E}}[f^2]} \Label{GDR}\\
{\le} & \min_{m}
\Delta_{2}(m',m|P_{AE}|P_E),\Label{GDR2}
\end{align}
where $f$ is a function taking values in $[0,1]$ and
\begin{align}
\Delta_{2}(m',m|P_{AE}|P_E)
:=
P_{AE}
\Big\{ -\log \frac{P_{AE}(a,e)}{P_E(e)} \le  m \Big \}
+
\frac{1}{2}e^{m'/2}
\sqrt{\frac{(P_{AE})^2}{P_E}
\Big\{ -\log \frac{P_{AE}(a,e)}{P_E(e)} >  m \Big \}}.
\end{align}
That is,
\begin{align}
\ell_{2}^{\varepsilon}(P_{AE}|P_E) \ge
\max\{m'|
\exists m \hbox{ such that }
\Delta_{2}(m',m|P_{AE}|P_E)
 \le \varepsilon\} 
\Label{GFT7}.
\end{align}
\hfill $\square$\end{lemma}

\begin{IEEEproof}
While we have $
\Delta_{2}(m'|P_{AE}|P_E)
=\min_{Q \in \bar{{\cal P}}({\cal A} \times {\cal E})} 
d(Q,P_{AE})+ 2^{(m'-H_2(Q|P_E))/2}$,
we can restrict $Q$ to a measure of the form $ P_{AE}f$.
In this case, $e^{-H_2(Q|P_E)}=\sum_{a,e}P_{AE}(a,e)^2f(a,e)^2P_E(e)$.
Hence, we obtain \eqref{GDR}.
We restrict the function $f$ to be a test function with support  
$\Big\{ -\log \frac{P_{AE}(a,e)}{P_E(e)} >  m \Big \}$.
Then, we obtain the inequality \eqref{GDR2}.
\end{IEEEproof}

The relation \eqref{GS3} follows from the following Lemma \ref{L9-B}.

\begin{lemma}\Label{L9-B}
For $P_{AE}\in {\cal P}({\cal A}\times {\cal E})$, we have
\begin{align}
{\ell}_{2}^\varepsilon(P_{AE}^n|P_E^n)
\ge nH(P_{AE}|P_E) +\sqrt{n}\sqrt{V(P_{AE}\|P_E)}\Phi^{-1}(\varepsilon)-\frac{1}{2}\log n+ F_{3}^\varepsilon(P_{AE}\|P_E)+O(\frac{1}{\sqrt{n}}).\Label{E22}
\end{align}
%where
%\begin{align}
%F_{3}^\varepsilon(P_{AE}\|P_E):=&
%\sqrt{V(P_{AE}\|P_E)} \Big(\frac{\kappa(P_{AE}\|P_E)(\Phi^{-1}(\varepsilon)^2-1)}{6 }-
%e^{ \frac{1}{2V(P_{AE}\|P_E)}+ \Phi^{-1}(\varepsilon)^2} \Big)+2 (\log 2-2) \\
%&- \log V(P_{AE}\|P_E) - \frac{1}{2}(\frac{1}{V(P_{AE}\|P_E)} +3 \Phi^{-1}(\varepsilon)^2).
%\end{align}
\hfill $\square$\end{lemma}

%$:= E \Big[ \Big(\frac{X-E[X]}{\sqrt{V[X]}}\Big)^3\Big]$,

\begin{IEEEproof}
We choose
$m:=nH(P_{AE}|P_E) +\sqrt{n}\sqrt{V(P_{AE}\|P_E)}B_1+ \sqrt{V(P_{AE}\|P_E)} B_2$
and $m':=m+2\log 2 -\frac{1}{2}\log (2\pi n) + B_3$ with $B_1=\Phi^{-1}(\varepsilon)$.
We apply Lemma \ref{L2} to the case with 
the measure $\frac{P_{AE}^2}{P_E} $ on ${\cal A} \times {\cal E}$, 
$X=-\log \frac{P_{AE}}{P_E} $, 
$R= \frac{m}{n}$,
$s_0=1$, and $R_0=H(P_{AE}|P_E)$.
Then, we obtain
\begin{align}
&\frac{m'}{2} +\frac{1}{2}\log{\frac{(P_{AE}^n)^2}{P_E^n}
\Big\{ -\log \frac{P_{AE}^n(a,e)}{P_E^n(e)} >  m \Big \}}\nonumber \\
=&\frac{m'}{2} +\frac{1}{2} \Big(-m -\frac{1}{2}\log (2\pi n) 
-\frac{1}{2}((\log V(P_{AE}\|P_E))+B_1^2)+v(d)+ O(\frac{1}{\sqrt{n}}) \Big)\nonumber \\
=&\log 2 -\frac{1}{2}\log (2\pi n) 
-\frac{1}{4}((\log V(P_{AE}\|P_E))+B_1^2)
+\frac{v(d)}{2}+\frac{B_3}{2}+ O(\frac{1}{\sqrt{n}}) .
\Label{E19}
\end{align}
Combining \eqref{E12} and \eqref{E19},
we have
\begin{align}
&
\Delta_{2}(m',m|P_{AE}^n|P_E^n)\nonumber \\
{=}& 
\Phi(B_1) 
+\frac{1}{\sqrt{2\pi n}}e^{- B_1^2/2}
(\frac{\kappa(P_{AE}\|P_E)(B_1^2-1)}{6 }
-B_2)+
\frac{1}{\sqrt{2\pi n} V(P_{AE}\|P_E)^{1/4}}e^{-\frac{1}{4}B_1^2+\frac{v(d)}{2}+\frac{1}{2}B_3}
+O(\frac{1}{n}) .\Label{E20}
\end{align}

The relation
\begin{align}
e^{- B_1^2/2}
(B_2-\frac{\kappa(P_{AE}\|P_E)(B_1^2-1)}{6 })
+\frac{1}{V(P_{AE}\|P_E)^{1/4}} e^{-\frac{1}{4}B_1^2+\frac{v(d)}{2}+\frac{1}{2}B_3}=0
\end{align}
holds if and only if 
\begin{align}
B_2=B_{2,2}(B_3)
:=\frac{\kappa(P_{AE}\|P_E)(B_1^2-1)}{6 }
- \frac{1}{V(P_{AE}\|P_E)^{1/4}} e^{\frac{1}{4} B_1^2+\frac{v(d)}{2}+\frac{1}{2}B_3}.\Label{E15B}
\end{align}
Lemma \ref{L3} implies that
\begin{align}
&\max_{B_3} \sqrt{V(P_{AE}\|P_E)} B_{2,2}(B_3)+B_3\nonumber \\
=&\max_{B_3}\sqrt{V(P_{AE}\|P_E)}
(\frac{\kappa(P_{AE}\|P_E)(B_1^2-1)}{6 }
-\frac{1}{V(P_{AE}\|P_E)^{1/4}}  e^{\frac{1}{4} B_1^2+\frac{v(d)}{2}+\frac{1}{2}B_3})+B_3 \nonumber \\
=&\sqrt{V(P_{AE}\|P_E)}\frac{\kappa(P_{AE}\|P_E)(B_1^2-1)}{6 }
+2\log 2 -2 -\frac{1}{2} \log V(P_{AE}\|P_E)
- \frac{1}{2}B_1^2-v(d).\Label{E21}
\end{align}
Due to the combination of \eqref{E20} and \eqref{E21} yields \eqref{E22},
when $B_2$ is chosen in \eqref{E15B} and $B_3$ is chosen to achieve the maximum in \eqref{E21}, 
the value $m'$ equals to the RHS of \eqref{E22}
because the sum of the RHS of \eqref{E21} and $2\log 2 -\frac{1}{2}\log (2\pi )$
equals $F_{3}^\varepsilon(P_{AE}\|P_E)$.
Hence,
$\max\{m'|
\exists m \hbox{ such that }
\Delta_{2}(m',m|P_{AE}^n|P_E^n)
 \le \varepsilon\} 
$ equals the RHS of \eqref{E22}.
Using Lemma \ref{L5X}, we obtain \eqref{E22}.
\end{IEEEproof}

\section{Proof of Theorems \ref{T3} and \ref{T4}}\Label{S9}
\subsection{Proof of Theorem \ref{T3}}\Label{S9-A}
In this section, we abbreviate $d_{(P\|Q)} $ to $d$.
We choose $m:=nD(P\|Q) +\sqrt{n}\sqrt{V(P\|Q)}B_1+ \sqrt{V(P\|Q)} B_2$ with $B_1= \Phi^{-1}(\varepsilon)$.
We apply Proposition \ref{L1} to the case with $X=\log \frac{P}{Q} $ and the distribution $P$.
Then, we obtain
\begin{align}
&P^n
\Big\{ \log \frac{P^n(a)}{Q^n(a)} \le  m \Big \}
\nonumber \\
=& \Phi(B_1) -\varphi(B_1)\frac{\kappa(P\|Q)(B_1^2-1)}{6 \sqrt{n}}
+\varphi(B_1) \frac{B_2}{\sqrt{n}}
+O(\frac{1}{n}) \nonumber \\
=& \varepsilon
+\frac{1}{\sqrt{2\pi n}}e^{- B_1^2/2}
(B_2-\frac{\kappa(P\|Q)(B_1^2-1)}{6 })
+O(\frac{1}{n}) \Label{E22-B}.
\end{align}
The relation $P^n \Big\{ \log \frac{P^n(a)}{Q^n(a)} \le  m \Big \}
=\varepsilon+ O(\frac{1}{n}) $ holds if and only if
\begin{align}
B_2= \frac{\kappa(P\|Q)(B_1^2-1)}{6 }.
\Label{E26}
\end{align}
Next, we apply Lemma \ref{L2} to the case with 
the measure $Q $ on ${\cal A}$, 
$X=\log \frac{P}{Q} $, 
$R=\frac{m}{n}$,
$s_0=1$, and $R_0=D(P\|Q)$.
Then, we obtain
\begin{align}
&
\log \Big[ Q^n \Big\{ \log \frac{P^n(a)}{Q^n(a)} >  m \Big \} \Big]
\nonumber \\
=&
-m -\frac{1}{2}\log (2 \pi V(P\|Q) n) - \frac{1}{2} B_1^2+v(d) +O(\frac{1}{\sqrt{n}}).
\Label{E23}
\end{align}
The combination of \eqref{E23} and \eqref{E26} yields \eqref{E27}.
%\end{IEEEproof}

\subsection{Proof of Theorem \ref{T4}}\Label{S9-B}
%\begin{IEEEproof}
Define
\begin{align}
\Delta_{DT}(m',m|P\|Q):=
P\Big\{ \log \frac{P(a)}{Q(a)} \le  m \Big \}
+e^{m'}
\Big[ Q \Big\{ \log \frac{P(a)}{Q(a)} >  m \Big \} \Big].
\end{align}
We find that
\begin{align}
\Delta_{DT}(m'|P\|Q)= \min_{m}
\Delta_{DT}(m',m|P\|Q).
\end{align}
Hence, we have
\begin{align}
D_{DT}^\varepsilon(P\|Q)=
\max\{m'|
\exists m \hbox{ such that }
\Delta_{DT}(m',m|P\|Q)
 \le \varepsilon\} .\Label{L5Z}
\end{align}
%\begin{align}
%&\Delta_{DT}(m'|P^n\|Q^n)-\varepsilon= \min_{m}
%P^n\Big\{ \log \frac{P^n(a)}{Q^n(a)} \le  m \Big \}-\varepsilon
%+e^{m'}\Big[ Q^n \Big\{ \log \frac{P^n(a)}{Q^n(a)} >  m \Big \} \Big].
%\end{align}
Due to \eqref{E23}, 
the second term has the order $1/\sqrt{n}$ if and only 
we have $m'=m+B_3$.
We choose $m$ in the same way as the proof of Theorem \ref{T3}.
Combining \eqref{E22-B} and \eqref{E23}, we have
\begin{align}
&
\Delta_{DT}(m',m|P^n\|Q^n) \nonumber \\
%P^n \Big\{ \log \frac{P^n(a)}{Q^n(a)} \le  m \Big \}
%+e^{m'} \Big[ Q^n \Big\{ \log \frac{P^n(a)}{Q^n(a)} >  m \Big \} \Big] \\
=& \varepsilon
+\frac{1}{\sqrt{2\pi n}}e^{- B_1^2/2}
(B_2-\frac{\kappa(P\|Q)(B_1^2-1)}{6 }
)
+\frac{1}{\sqrt{2\pi V(P\|Q) n}}
e^{B_3-\frac{1}{2} B_1^2+v(d)}
+O(\frac{1}{n}) .
\end{align}
 Hence,
 \begin{align}
(B_2-\frac{\kappa(P\|Q)(B_1^2-1)}{6 })
+\frac{1}{V(P\|Q)^{1/2}}e^{B_3+v(d)}
 =0
 \end{align}
 if and only if
 \begin{align}
B_2= 
\frac{\kappa(P\|Q)(B_1^2-1)}{6 }
-\frac{1}{V(P\|Q)^{1/2}} e^{B_3+v(d)}.\Label{E15C}
 \end{align}
 Lemma \ref{L3} implies that
\begin{align}
&\max_{B_3}\sqrt{V(P\|Q)}
(\frac{\kappa(P\|Q)(B_1^2-1)}{6 }
-\frac{1}{V(P\|Q)^{1/2}}e^{B_3+ v(d)})+B_3 \nonumber \\
=&\sqrt{V(P\|Q)}\frac{\kappa(P\|Q)(B_1^2-1)}{6 } 
-\frac{1}{2}\log V(P\|Q) -v(d)-1.\Label{E29}
\end{align}
%Due to the combination of \eqref{E20} and \eqref{E21} yields \eqref{E22},
When $B_2$ is chosen in \eqref{E15C} and $B_3$ is chosen to achieve the maximum in \eqref{E29}, 
the value $m'$ equals to the RHS of \eqref{E27B}
because the RHS of \eqref{E29} equals $F_{5}^\varepsilon(P\|Q)$.
Hence,
$
\max\{m'|
\exists m \hbox{ such that }
\Delta_{DT}(m',m|P^n\|Q^n)
 \le \varepsilon\} 
$ equals the RHS of \eqref{E27B}.
Using \eqref{L5Z}, we obtain \eqref{E27B}.
%\end{IEEEproof}

%%%%%%%%%%% Discussions %%%%%%%%%%%%%%%%%%%%%%%
\section{Conclusions}\Label{S10}

\begin{savenotes}
\begin{table*}
%\begin{table}[htb]
%\begin{center}
  \caption{Summary of obtained results.}
\begin{center}
  \begin{tabular}{|l|c|c|c|} 
\hline
 {Setting} & Direct & Converse  & Matched Order \\ \hline
Secure Random Num. Gen.  &  \multicolumn{2}{|c|}{Theorem \ref{T2} and \eqref{LGT2}}   & $O(\sqrt{n})$ \\ \hline
Binary hypothesis testing  & \multicolumn{2}{|c|}{Theorems \ref{T3} and \ref{T4}}   & $O(1)$ \\ \hline
Source coding with side info. & \multicolumn{2}{|c|}{Theorem \ref{T6} }   & $O(\sqrt{n})$ \\ \hline
Channel coding  & Theorems \ref{T8} and \ref{T9} & Theorem \ref{T10}  & $O(\sqrt{n})$ \\ \hline
Wire-tap channel coding  & Theorems \ref{T11} and \ref{T12} & Theorem \ref{T13}   & $O({n})$ \\ \hline
  \end{tabular}
\end{center}

\vspace{2ex}

Matched order means the lowest order, in which, the upper bound and lower bound match each other.
\Label{table1}
\end{table*}
\end{savenotes}

In this paper, we have made semi-finite length analysis for upper and lower bounds
for various problems, secure random number generation,
simple binary hypothesis testing,
fixed-length source coding with and without side information,
channel coding with conditional additive channel, and
wire-tap channel coding with conditional additive and degraded channel.
Obtained results are summarized in Table \ref{table1}.
Unfortunately, we could not discuss 
the random coding union (RCU) bound
because it requires more complicated evaluation.
Since the RCU bound is better than the DT bound,
higher order expansion of the RCU bound is an interesting future problem.

\section*{Acknowledgment}
The author was supported in part by JSPS Grant-in-Aid for Scientific Research (B) No.16KT0017 and for Scientific Research (A) No.17H01280,
and Kayamori Foundation of Informational Science Advancement.
He is grateful to 
Professor \'{A}ngeles V\'{a}zquez-Castro,
Professor Vincent Y. F. Tan,
and Dr. Wei Yang 
for helpful discussions.

%%%%%%%%%%% Appendix %%%%%%%%%%%%%%%%%%%%%%%%
\appendices

%%%%%
\section{Proof of Proposition \ref{lemma:monotonicity}}
\label{appendix:lemma:monotonicity}

Let $\tilde{P}_{SZ} \in {\cal B}^\varepsilon(P_{SZ})$ be such that 
\begin{eqnarray*}
H_{\min}^\varepsilon(P_{SZ}|R_E) = H_{\min}(\tilde{P}_{SZ}|R_E).
\end{eqnarray*}
Then, we define
\begin{eqnarray*}
\tilde{P}_{AE}(a,e) = 
\tilde{P}_{SZ}(f(x),z) \frac{P_{AE}(a,e)}{P_{SZ}(f(x),z)}.
\end{eqnarray*}
Then, we have
\begin{eqnarray*}
\lefteqn{ d(\tilde{P}_{AE},  P_{AE}) } \\
&=& \frac{1}{2} \sum_{x,z} |\tilde{P}_{AE}(a,e) - P_{AE}(a,e) | \\
&=& \frac{1}{2} \sum_{s,z} \sum_{x \in f^{-1}(s)} \frac{P_{AE}(a,e)}{P_{SZ}(s,z)} |\tilde{P}_{SZ}(s,z) - P_{SZ}(s,z) | \\
&=& \frac{1}{2} \sum_{s,z}  |\tilde{P}_{SZ}(s,z) - P_{SZ}(s,z) | \\
&=& d(\tilde{P}_{SZ},  P_{SZ} ) \\
&\le& \varepsilon.
\end{eqnarray*}
Thus, we have $\tilde{P}_{AE} \in  {\cal B}^\varepsilon(P_{AE})$. Furthermore, by the construction of $\tilde{P}_{AE}$, we have
$\tilde{P}_{AE}(a,e) \le \tilde{P}_{SZ}(f(x),z)$ for every $(a,e)$. Thus, we have
\begin{eqnarray*}
H_{\min}^\varepsilon(P_{SZ}|R_E) &=& H_{\min}(\tilde{P}_{SZ}|R_E) \\
&\le& H_{\min}(\tilde{P}_{AE}|R_E) \\
&\le& H_{\min}^\varepsilon(P_{AE}|R_E).
\end{eqnarray*}
\qed

\if0
\begin{remark}
When Eve's side-information is the quantum density operator instead of
the random variable, the monotonicity of the smooth minimum entropy was
proved in \cite[Proposition 3]{tomamichel:12}, where the smoothing
is evaluated by the so-called purified distance instead of the trace distance.
For the quantum setting and the trace distance, it is not clear whether the monotonicity holds
or not because we cannot apply Uhlmann's theorem to the trace distance directly.
\end{remark}
\fi

\section{Proof of Proposition \ref{T1}}\Label{A2}
Since the first inequality in \eqref{LGT} follows from \eqref{LBD},
we show the second inequality in \eqref{LGT}.
Then, we define
\begin{align*}
\bar{H}_{\min}^\varepsilon(P_{AE}|R_E) &:= \max_{Q_{AE} \in \bar{{\cal B}}^\varepsilon(P_{AE})} H_{\min}(Q_{AE}|R_E), 
%\bar{\delta}_{\min}(m | P_{AE}|R_E)&:= \min_{Q_{AE} \in \bar{{\cal P}}({\cal A} \times {\cal E})}
%\{ d(Q_{AE}, P_{AE})|  H_{\min}(Q_{AE}|R_E) \le m \}
\end{align*}
where 
\begin{eqnarray*}
\bar{{\cal B}}^\varepsilon(P_{AE}) = \left\{ Q_{AE} \in \bar{{\cal P}}({\cal A} \times {\cal E}) : d(P_{AE}, Q_{AE}) \le \varepsilon \right\}.
\end{eqnarray*}

The following is a key lemma to derive 
every lower bound of $\ell(P_{AE},\varepsilon)$.
\begin{lemma}[Leftover Hash:\cite{renner:05b},\cite{bennett:95},\cite{HILL}]
\label{lemma:left-over}
Let $F$ be the uniform random variable on a set of universal 2 hash family ${\cal F}$. Then, 
for $P_{AE} \in \bar{{\cal P}}({\cal A} \times {\cal E})$ and
$R_E \in {\cal P}({\cal E})$, we have\footnote{Technically,
$R_E$ must be such that $\rom{supp}(P_E) \subset \rom{supp}(R_E)$.}
\begin{eqnarray*}
\mathbb{E}_F[d(F|P_{AE}) ] \le \frac{1}{2} \sqrt{|{\cal S}| 2^{- H_2(P_{AE}|R_E)}}.
\end{eqnarray*}
\hfill $\square$\end{lemma}

Furthermore, since
\begin{eqnarray*}
d(P_{AE}|f) \le 2 \varepsilon + d(\bar{P}_{AE}|f)
\end{eqnarray*}
holds for $\bar{P}_{AE} \in \bar{{\cal B}}^\varepsilon(P_{AE})$ by the triangular inequality, we have the following.
\begin{corollary}[\protect{\cite[Corollary 2]{ISIT2013}}]
\Label{corollary:smooth-entropy-bound}
For $P_{AE} \in {\cal P}({\cal A} \times {\cal E})$ and $R_E \in {\cal P}({\cal E})$, we have
\begin{eqnarray*}
\mathbb{E}_F[d(F|P_{AE}) ] \le 
2 \varepsilon + \frac{1}{2} \sqrt{|{\cal S}| 2^{- \bar{H}_{\min}^\varepsilon(P_{AE}|R_E)}}.
\end{eqnarray*}
\end{corollary}
Corollary \ref{corollary:smooth-entropy-bound} implies the second inequality in \eqref{LGT}.

\section{Proof of Second inequality in Proposition \ref{P8}}\Label{A4}
Given the memory set ${\cal M} $ with the cardinality $\mathsf{M}$,
we randomly choose the encoder $F$ such that  
\begin{align}
\mathbb{P} \{  F(x)=F(x')\}\le \frac{1}{\mathsf{M}}.
\end{align}
Given a encoder $f: {\cal X} \to {\cal M}$, 
we define decoder $g_f $ as follows.
Given $m \in {\cal M}$ and $y \in {\cal Y}$,
we decide $g_f(m,y)$ to be an element $x \in {\cal X}$ to satisfy that $y \in Q_x$
where $Q_{x}:=\{y| P_{XY}(x,y) \ge \frac{1}{\mathsf{M}}  P_Y(y)\}$.
If no element $x \in {\cal X}$ satisfies this condition,
we decide $g_f(m,y)$ to be an arbitrary element of ${\cal X}$.
In this code, the decoding error probability is upper bounded by
\begin{align}
&\sum_{x} P_X(x)P_{Y|X=x}Q_x^c
+
\sum_{x} P_X(x) \sum_{x'(\neq x) \in {\cal X}| f(x)=f(x')} P_{Y|X=x}Q_{x'}\Label{CRT}
\end{align}
The average of the second term with respect to the choice of $f$
is evaluated as follows.
\begin{align}
& \mathbb{P_F}
\sum_{x} P_X(x) \sum_{x' (\neq x) \in {\cal X}| F(x)=F(x')} P_{Y|X=x}Q_{x'}  \nonumber\\
\le
&
\sum_{x} P_X(x) \sum_{x' (\neq x) \in {\cal X}}\frac{1}{\mathsf{M}} 
P_{Y|X=x}Q_{x'} 
\le
\sum_{x} P_X(x) \sum_{x'  \in {\cal X}}\frac{1}{\mathsf{M}} 
P_{Y|X=x}Q_{x'} \nonumber\\
=
& 
\sum_{x'  \in {\cal X}}\frac{1}{\mathsf{M}} P_{Y}Q_{x'} 
=
\frac{1}{\mathsf{M}} P_{Y}\times I 
\Big\{(x,y)\Big| P_{XY}(x,y) \ge \frac{1}{\mathsf{M}}  P_Y(y)\Big\}.
\end{align}
Hence, the average of \eqref{CRT} is upper bounded by
\begin{align}
P_{XY}
\Big\{(x,y)\Big| P_{XY}(x,y) < \frac{1}{\mathsf{M}}  P_Y(y)\Big\}
+
\frac{1}{\mathsf{M}} P_{Y}\times I 
\Big\{(x,y)\Big| P_{XY}(x,y) \ge \frac{1}{\mathsf{M}}  P_Y(y)\Big\}
=\Delta_{DT} (P_{XY}\|P_Y).\Label{LDT}
\end{align}
This evaluation 
with yields the second inequality of \eqref{HUT}.

\begin{remark}
The paper \cite[Theorem 7]{tomamichel:12}
derived the upper bound 
\begin{align}
(1+c)
P_{XY}
\Big\{(x,y)\Big| P_{XY}(x,y) < \frac{1}{\mathsf{M}}  P_Y(y)\Big\}
+
\frac{(c+1)^2}{c}\frac{1}{\mathsf{M}} P_{Y}\times I 
\Big\{(x,y)\Big| P_{XY}(x,y) \ge \frac{1}{\mathsf{M}}  P_Y(y)\Big\}\Label{MMR}
\end{align}
in their proof in the quantum setting.
Since it considers the quantum setting, 
the coefficients in their upper bound are   
$(1+c)$ and $\frac{(c+1)^2}{c}$ due to the use of 
Hayashi-Nagaoka inequality to handles the non-commutativity.
In the commutative setting,
we can replace them by $1$.
Hence, the upper bound \eqref{MMR} equals the upper bound \eqref{LDT}.
\end{remark}

\end{document}